\documentclass[aps,prd,eprint,numerical,showpacs,showkeys,10pt]{revtex4-1}

\usepackage{colortbl}
\usepackage{longtable}
\usepackage{amsmath}
\usepackage{amssymb}
\usepackage{amsfonts}
\usepackage{graphicx}
\usepackage{dcolumn}
\usepackage{bm}
\usepackage{hyperref}
\hypersetup{colorlinks=false}

\begin{document}

\title[Scalar quasibound states in the 4DEGB(A)dS]{Quasibound states of scalar fields in the consistent 4D Einstein-Gauss-Bonnet-(Anti-)de Sitter gravity}

\date{\today}

\author{H. S. Vieira}
\email{Corresponding author: horacio.santana.vieira@hotmail.com; horacio.santana-vieira@tat.uni-tuebingen.de}
\affiliation{Theoretical Astrophysics, Institute for Astronomy and Astrophysics, University of T\"{u}bingen, 72076 T\"{u}bingen, Germany}

\author{V. B. Bezerra}
\email{valdir@fisica.ufpb.br}
\affiliation{Departamento de F\'{i}sica, Universidade Federal da Para\'{i}ba, Caixa Postal 5008, CEP 58051-970, Jo\~{a}o Pessoa, PB, Brazil}

\author{C. R. Muniz}
\email{celio.muniz@uece.br}
\affiliation{Grupo de F\'{i}sica Te\'{o}rica (GFT), Universidade Estadual do Cear\'{a}, UECE-FECLI, Iguatu, CE, Brazil}

\author{M. S. Cunha}
\email{marcony.cunha@uece.br}
\affiliation{Grupo de F\'isica Te\'orica (GFT), Centro de Ci\^encias e Tecnologia, Universidade Estadual do Cear\'a, CEP 60714-903, Fortaleza, CE, Brazil}

\begin{abstract}
We examine the interaction between massless scalar fields and the gravitational field generated by a black hole solution that was recently obtained in the consistent well-defined 4-dimensional Einstein-Gauss-Bonnet gravity with a cosmological constant. In order to do this, we calculate quasibound state frequencies of scalar fields for the spherically symmetric black hole in the consistent 4-dimensional Einstein-Gauss-Bonnet-de Sitter and Anti-de Sitter theories. The expression for the quasibound states is obtained by using the polynomial condition associated to the Heun functions, and their values are overdamped. We also demonstrate the stability of the systems.
\end{abstract}



\preprint{Preprint submitted to The European Physical Journal C}
%\preprint{AIP/123-QED}

\maketitle

%\begin{quotation}
%...
%\end{quotation}

%
%%%%%%%%%%%%%%%%%%%%%%%%%%%%%%%%%%%%%%%%%%%%%%%%%%%%%%%%%%%%%%%%%%%%%%%%%%%%%%%%%%%%%%%%%%%%%% Introduction
%
\section{Introduction}\label{Introduction}
Nowadays, some modern experiments \cite{PhysRevLett.116.061102}, which were designed to observe a transient gravitational-wave signal, may also provide an unprecedented opportunity to test Einstein's general theory of relativity at the strong-field regime, as for example, by studying the two-body motion of a compact-object binary \cite{PhysRevLett.116.221101}, or a binary black hole merger as well \cite{Universe.7.497}. This does not exclude alternative/modified theories of gravity, quite the opposite, in order to interpret some of these data, it could be used, in principle, such theories, including the ones with higher order curvature corrections to the Einstein term, such as the Einstein-Gauss-Bonnet (EGB) gravity, or even the Lovelock theory of gravity \cite{PhysRevD.98.024042,IntJModPhysD.26.1730001}, since some fundamental questions, including the issues related to the quantum gravity phenomenon, cannot be solved with the Einstein's theory.

In order to deal with the higher curvature corrections that appears in the standard Einstein's theory (see Ref.~\cite{PhysRevLett.125.149001} and references therein), it was developed the EGB theory, which is quadratic in the curvature and one of the most promising approaches among the alternative/modified theories of gravity. In the case when the Gauss-Bonnet (GB) term is coupled to a matter field, the EGB theory leads to non-trivial corrections of the equation of motion. The so-called well-defined and consistent theory of 4-dimensional EGB (4DEGB) theory of gravity was developed by Aoki-Gorji-Mukohyama (AGM) \cite{PhysLettB.810.135843,JCAP.09.014,JCAP.05.E01}, which uses the Arnowitt-Deser-Misner (ADM) decomposition \cite{PhysRev.116.1322} to construct the Hamiltonian in such a way that this (new) theory has not infinite coupling. It is worth drawing attention to the comments made by Arrechea \textit{et al.} \cite{PhysRevLett.125.149002,ChinPhysC.45.013107} about a well-defined and consistent 4DEGB theory of gravity.

Among the papers dealing with the 4DEGB theory of gravity, we can mention the one which studies the quasibound states of the scalar, electromagnetic and Dirac test fields in a spherically symmetric asymptotically flat black holes \cite{arXiv:2107.02065}, as well as the work which investigates the quasinormal modes of these test fields in the 4D Einstein-Gauss-Bonnet-(anti)de Sitter (4DEGBAdS) gavity \cite{AnnPhys.427.168425}.

In the present paper, we extend these results by studying the quasibound states for the conformally coupled massless scalar field in both 4D Einstein-Gauss-Bonnet-de Sitter (4DEGBdS) and 4D Einstein-Gauss-Bonnet-Anti-de Sitter (4DEGBAdS) backgrounds. These calculations are carried out in the stability region of the GB coupling constant \cite{EurPhysJC.80.1049}. It is worth emphasizing that the cosmological constant should be, in principle, a negligible factor in the studies of black holes, since its current value is quite small and hence it does not play any role in the black hole dynamics. However, an effective cosmological constant can appear when regions with high curvature in alternative/modified theories of gravity are considered. In this case, such term could, in principle, describe the interaction of the black hole with a dark energy component of the universe. Therefore, it is interesting to investigate the role played by the cosmological constant within the 4DEGB theory.

In this work, we will calculate the spectrum of quasibound states, as well as the corresponding angular and radial wave eigenfunctions, for massless scalar particles in the 4DEGB black hole solutions with a cosmological constant by using the polynomial condition of the Heun functions, the so-called Vieira-Bezerra-Kokkotas (VBK) approach (for details, see Refs. \cite{AnnPhys.373.28,PhysRevD.104.024035}). We show that the quasibound states depend on the GB coupling constant, $a$, as well as on the cosmological constant, $\Lambda$. We also investigate the stability of the systems.

The quasibound states \cite{LettNuovoCim.15.257,RomJPhys.38.729} are a kind of wave phenomena occurring near the black hole exterior event horizon. They are localized in the black hole potential well, which means that there exist a flux of particles crossing into the black hole. Thus, the spectrum of quasibound states is constituted by complex frequencies, which can be expressed as $\omega=\omega_{R}+i\omega_{I}$, where $\omega_{R}$ and $\omega_{I}$ are the real and imaginary parts, respectively. The real part of the resonant frequency is the oscillation frequency, while the imaginary part determines the stability of the system. Thus, the wave solution is said to be stable when the imaginary part of the resonant frequency is negative ($\omega_{I} < 0$), which signalizes the existence of a decay rate with the time. Otherwise, the wave solution is unstable when the imaginary part of the resonant frequency is positive ($\omega_{I} > 0$)and therefore, contrarily to the previous case, there is a growth rate with the time.

The paper is organized as follows. In Sec. \ref{4DEGB}, we introduce the general metric corresponding to the 4DEGB black hole spacetimes with a cosmological constant. In Sec. \ref{KGE}, we separate the conformally coupled massless Klein-Gordon equation in the background under consideration. In Sec. \ref{4DEGBdS}, we discuss the aforementioned metric for the case of a positive cosmological constant (4DEGBdS black hole) and find an exact solution for the radial equation in this background. Then, we obtain an expression for the quasibound states. In Sec. \ref{4DEGBAdS}, we repeat this analysis for the case of a negative cosmological constant (4DEGBAdS black hole). Finally, in Sec. \ref{Conclusions}, we summarize the obtained results. Here we adopt the natural units where $G \equiv c \equiv \hbar \equiv 1$.
%
%%%%%%%%%%%%%%%%%%%%%%%%%%%%%%%%%%%%%%%%%%%%%%%%%%%%%%%%%%%%%%%%%%%%%%%%%%%%%%%%%%%%%%%%%%%%%% The consistent well-defined 4-dimensional Einstein-Gauss-Bonnet gravity with a cosmological constant
%
\section{The consistent well-defined 4-dimensional Einstein-Gauss-Bonnet gravity with a cosmological constant}\label{4DEGB}
In the studies of black hole radiation, which means the investigation of emission/transmission/reflection of any kind of quasispectrum, it is crucial that the black hole solutions should be solutions of both truly 4-dimensional AGM (4DAGM) \cite{PhysLettB.810.135843,JCAP.09.014,JCAP.05.E01} and extra scalar degrees of freedom \cite{JHEP.04.082,PhysRevD.88.024006,JCAP.01.017,PhysRevD.102.024025,JCAP.07.013,PhysLettB.808.135657,PhysLettB.809.135717,PhysRevLett.124.081301} theories. Thus, we can call it as a consistent well-defined 4-dimensional Einstein-Gauss-Bonnet theory of gravity, which includes the black hole solutions with a cosmological constant.

Then, by taking into account all of the aforementioned approaches, we will briefly review the basic ideas behind it in order to present the 4DEGB black hole solutions with a cosmological constant.

Let us start by stating that the consistent well-defined 4DEGB gravity is based on the ADM formalism, where the 4D metric is defined by
\begin{equation}
ds^{2}=-\mathcal{N}^{2}dt^{2}+\gamma_{ij}(dx^{i}+\mathcal{N}^{i}dt)(dx^{j}+\mathcal{N}^{j}dt),
\label{eq:ADM_metric_4DEGB}
\end{equation}
with $\mathcal{N}$ being the so-called lapse function, $\mathcal{N}^{i}$ is a shift vector, and $\gamma_{ij}$ is the spatial metric. In a 4D spacetime, the following gauge condition is valid:
\begin{equation}
^{3}\mathcal{G} = \sqrt{\gamma}D_{k}D^{k}(\pi^{ij}\gamma_{ij}/\sqrt{\gamma}) \approx 0,
\label{eq:gauge_4DEGB}
\end{equation}
where $D_{k}$ is the covariant derivative and $\pi^{ij}$ is the canonical momentum conjugate to $\gamma_{ij}$. Then, the gravitational action, $S$, is defined as
\begin{equation}
S=\int dt\ d^{3}x\ \mathcal{N}\ \sqrt{\gamma}\ \mathcal{L}_{\rm 4DEGB},
\label{eq:action_4DEGBAdS}
\end{equation}
with
\begin{equation}
\mathcal{L}_{\rm 4DEGB}=\frac{1}{2\kappa^{2}}(-2\Lambda+\mathcal{K}_{ij}\mathcal{K}^{ij}-\mathcal{K}^{i}_{i}\mathcal{K}^{j}_{j}+\mathcal{R}+a\mathcal{R}^{2}_{4DGB}),
\label{eq:Lagrangian_4DEGBAdS}
\end{equation}
\begin{equation}
\mathcal{R}^{2}_{4DGB}=-\frac{4}{3}(8\mathcal{R}_{ij}\mathcal{R}^{ij}-4\mathcal{R}_{ij}\mathcal{M}^{ij}-\mathcal{M}_{ij}\mathcal{M}^{ij})+\frac{1}{2}(8\mathcal{R}^{2}-4\mathcal{R}\mathcal{M}-\mathcal{M}^{2}),
\label{eq:Riemann_4DEGBAdS}
\end{equation}
\begin{equation}
\mathcal{M}_{ij}=\mathcal{R}_{ij}+\mathcal{K}^{k}_{k}\mathcal{K}_{ij}-\mathcal{K}_{ik}\mathcal{K}^{k}_{j},
\label{eq:M_tensor_4DEGBAdS}
\end{equation}
\begin{equation}
\mathcal{K}_{ij}=\frac{1}{2\mathcal{N}}(\dot{\gamma}_{ij}-2D_{(i}\mathcal{N}_{j)}-\gamma_{ij}D^{2}\lambda_{\rm GF}),
\label{eq:K_tensor_4DEGBAdS}
\end{equation}
where $\kappa$ is the gravitational coupling (strength) constant, $a$ is the (rescaled) GB coupling constant, $\mathcal{R}_{ij}$ is the Ricci tensor, $\mathcal{R}$ is the Ricci scalar, $\lambda_{\rm GF}$ is a gauge-fixing (GF) parameter, and $\mathcal{M}=\mathcal{M}^{i}_{i}$. Here, the dot denotes the derivative with respect to the time $t$. The final step is to choose the GF parameter such that $\lambda_{\rm GF}=0$ and appropriately rescale both the gravitational and GB coupling constants, as well as to take $M$ as the total mass centered at the origin of the system of coordinates. Therefore, an exact solution describing the 4DEGB black hole spacetime with a cosmological constant has the following form
\begin{equation}
ds^{2}=-f_{\pm}(r)\ dt^{2}+\frac{1}{f_{\pm}(r)}\ dr^{2}+r^{2}\ d\theta^{2}+r^{2}\sin^{2}\theta\ d\phi^{2},
\label{eq:metric_4DEGBAdS}
\end{equation}
where the metric function, $f_{\pm}(r)$, is given by
\begin{equation}
f_{\pm}(r)=1+\frac{r^{2}}{a}\Biggl[1 \pm \sqrt{1+2a\biggl(\frac{2M}{r^{3}}+\frac{\Lambda}{3}\biggr)}\ \Biggr].
\label{eq:metric_function_4DEGBAdS}
\end{equation}
The metric function $f_{+}(r)$ corresponds to an asymptotically de Sitter spacetime in the limit when $\Lambda \rightarrow 0$. On the other hand, the metric function $f_{-}(r)$ corresponds to an asymptotically flat spacetime when $\Lambda \rightarrow 0$. Here, we focus on the ``minus'' case and hence we set $f(r) \equiv f_{-}(r)$, such that the surface equation can be written as
\begin{equation}
f(r)=0=r^{4}-\frac{3r^{2}}{\Lambda}+\frac{6Mr}{\Lambda}-\frac{3a}{2\Lambda}.
\label{eq:surface_equation_4DEGBAdS}
\end{equation}
In general, Eq.~(\ref{eq:surface_equation_4DEGBAdS}) has four solutions, which can be complex numbers, as well as positive and negative real numbers. In the present work, we will consider only the real solutions, including the negative ones, which may describe the (Cauchy) ``interior'' cosmological horizon. It is worth emphasizing that we will obtain these solutions directly from Eq.~(\ref{eq:metric_function_4DEGBAdS}) by using the built-in symbol \textsl{Solve} on Wolfram Mathematica 12.3.

Next, we will consider massless scalar fields in the 4DEGB black hole spacetime with a cosmological constant given by Eq.~(\ref{eq:metric_4DEGBAdS}).
%
%%%%%%%%%%%%%%%%%%%%%%%%%%%%%%%%%%%%%%%%%%%%%%%%%%%%%%%%%%%%%%%%%%%%%%%%%%%%%%%%%%%%%%%%%%%%%% The Klein-Gordon equation
%
\section{The Klein-Gordon equation}\label{KGE}
We are interested in some basic characteristics of these 4DEGB black hole spacetime with a cosmological constant, in particular the ones related to their interaction with quantum fields, including the Hawking radiation, and classical scalar wave scattering such as the quasibound states (QBSs). To this end, we want to consider the conformally coupled massless scalar field as a probe, whose covariant equation of motion is given by
\begin{equation}
\frac{1}{\sqrt{-g}}\partial_{\mu}(g^{\mu\nu}\sqrt{-g}\partial_{\nu}\Phi)-\frac{1}{6}\mathcal{R}\Phi=0.
\label{eq:conformal_massless_KGE}
\end{equation}
In order to obtain solutions of the conformally coupled massless Klein-Gordon equation (\ref{eq:conformal_massless_KGE}), and due to stationarity and axisymmetry, we use the following separation ansatz
\begin{equation}
\Phi(t,r,\theta,\phi)=\mbox{e}^{-i \omega t}U(r)P(\theta)\mbox{e}^{i m \phi},
\label{eq:ansatz}
\end{equation}
where $U(r)$ is the radial function, $P(\theta)$ is the polar angle function, $m$ $(\in \mathbb{Z})$ is the magnetic or azimuthal quantum number, and $\omega$ is the frequency (energy, in the natural units). By substituting Eq.~(\ref{eq:ansatz}) into Eq.~(\ref{eq:conformal_massless_KGE}), we obtain two ordinary differential equations, namely,
\begin{equation}
\frac{d^{2} P(\theta)}{d \theta^{2}}+\frac{\cos\theta}{\sin\theta}\frac{d P(\theta)}{d \theta}+\biggl(\lambda-\frac{m^{2}}{\sin^{2}\theta}\biggr)P(\theta)=0,
\label{eq:angular_equation_4DEGB}
\end{equation}
\begin{equation}
\frac{d^{2} U(r)}{d r^{2}}+\biggl[\frac{2}{r}+\frac{1}{f(r)}\frac{d f(r)}{d r}\biggr]\frac{d U(r)}{d r}+\frac{1}{f^2(r)}\biggl[\omega^2-\frac{1}{6}Rf(r)-\frac{\lambda f(r)}{r^2}\biggr]U(r)=0,
\label{eq:radial_equation_4DEGB}
\end{equation}
where $\lambda$ is the separation constant and the prime denotes differentiation of the polar and radial functions with respect to $\vartheta$ and $r$, respectively. The Ricci scalar $\mathcal{R}$ can be written in terms of the metric function $f(r)$ as
\begin{equation}
\mathcal{R}=-\frac{1}{r^{2}}\biggl[r^{2}\frac{d^{2} f(r)}{d r^{2}}+4r\frac{d f(r)}{d r}+2f(r)-2\biggr].
\label{eq:Ricci_scalar}
\end{equation}
The final (parametrized) expression for the Ricci scalar $\mathcal{R}$ will depend on the parametrization adopted for the metric function $f(r)$.

The general solution of the polar equation (\ref{eq:angular_equation_4DEGB}) is given in terms of the associated Legendre functions $P(\theta)=P_{\nu}^{m}(\cos\theta)$ with general degree $\nu$ (which can be a complex number $\mathbb{C}$) and $m \geq 0$ $\in \mathbb{Z}$, such that $\lambda=\nu(\nu+1)$.

Now, let us turn our attention to the solution of the radial equation. Firstly, we will analyze the solutions of the surface equation $f(r)=0$, that is, the number of event horizons, and then write the metric function $f(r)$ in a more convenient form to write Eq.~(\ref{eq:radial_equation_4DEGB}) as a Heun-type equation that is useful to obtain exact solutions, which is useful to study the Hawking radiation and QBS spectra, which will be presented in the next sections.
%
%%%%%%%%%%%%%%%%%%%%%%%%%%%%%%%%%%%%%%%%%%%%%%%%%%%%%%%%%%%%%%%%%%%%%%%%%%%%%%%%%%%%%%%%%%%%%% 4DEGBdS
%
\section{The 4DEGB-de Sitter black hole spacetime}\label{4DEGBdS}
For a 4DEGBdS black hole spacetime, we fix $M=1/2$ and then calculate the black hole event horizons according to the values of the cosmological constant ($\Lambda > 0$), as well as for the GB coupling constant in the stability region $-0.20 \leq a \leq 0.20$ \cite{EurPhysJC.80.1049}.

\begin{figure}[p]
\centering
\includegraphics[width=1\columnwidth]{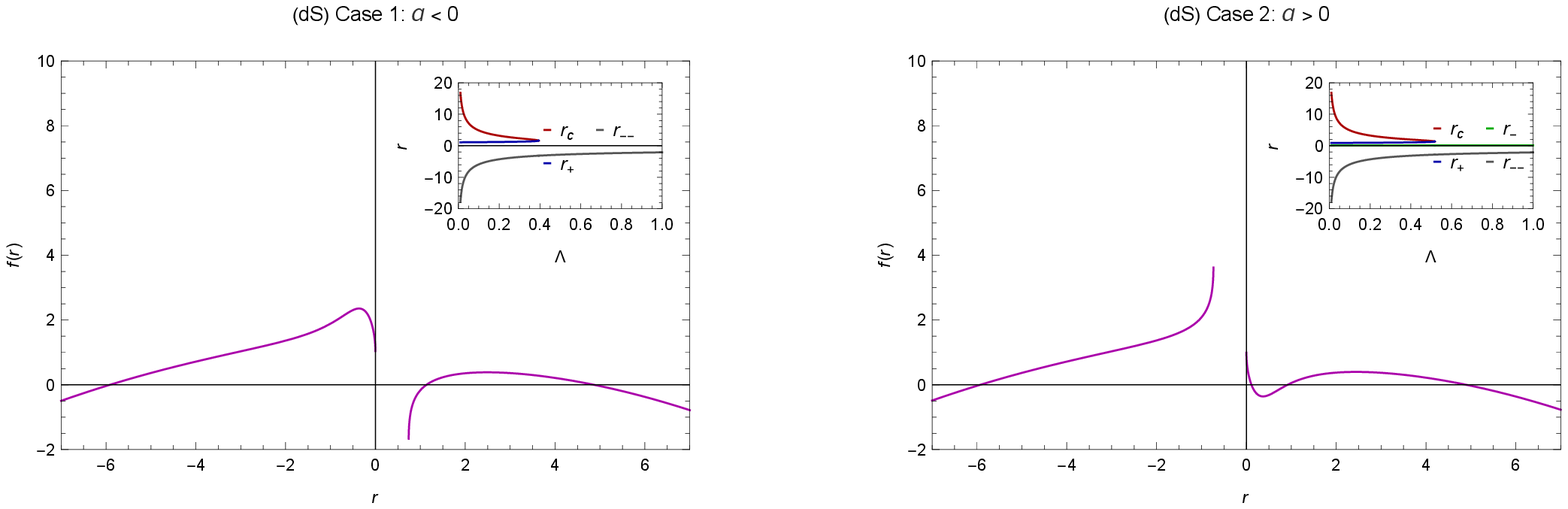}
\caption{The metric function $f(r)$ for $a=-0.20$ (left) and $a=0.20$ (right); $\Lambda=0.1$, $M=1/2$.}
\label{fig:Fig1_4DEGBdS}
\end{figure}

The behavior of the metric function $f(r)$ is shown in Fig.~\ref{fig:Fig1_4DEGBdS}. In addition, the horizons are shown in the inlay plots as functions of the cosmological constant $\Lambda$. We can see that the surface equation $f(r)=0$ has three real solutions when $a < 0$, otherwise, there are four real solutions when $a > 0$.

%
%%%%%%%%%%%%%%%%%%%%%%%%%%%%%%%%%%%%%%%%%%%%%%%%%%%%%%%%%%%%%%%%%%%%%%%%%%%%%%%%%%%%%%%%%%%%%% (dS) Case 1: a < 0
%
\subsection{(dS) Case 1: \texorpdfstring{$a < 0$}{a < 0}}
In this case, we adopt the following parametrization for the metric function,
\begin{equation}
f(r)=\frac{1}{r^{2}}(r-r_{c})(r-r_{+})(r-r_{--}),
\label{eq:metric_function_4DEGBdS_1}
\end{equation}
where $r_{c}$ is the (positive) ``exterior'' de Sitter cosmological horizon, $r_{+}$ is the (positive) ``exterior'' black hole event horizon, and $r_{--}$ is the (negative) ``interior'' de Sitter cosmological horizon. Then, with the metric function given by Eq.~(\ref{eq:metric_function_4DEGBdS_1}), we will show that Eq.~(\ref{eq:radial_equation_4DEGB}) is totally appropriate to study QBSs with purely ingoing boundary conditions at the exterior black hole event horizon and vanishing boundary conditions at infinity, since it is a (general) Heun-type equation \cite{Ronveaux:1995} with three finite regular singularities (associated to the three event horizons) and one regular singularity at (spatial) infinity.

Now, we follow the steps described in the VBK approach \cite{AnnPhys.373.28,PhysRevD.104.024035} to obtain the analytical solution of Eq.~(\ref{eq:radial_equation_4DEGB}) in the 4DEGBdS black hole spacetime with $a < 0$ (without the assumption of specific boundary conditions). First of all, we need to define a new radial coordinate, $x$, as
\begin{equation}
x=\frac{r_{--}-r}{r_{c}-r}\eta,
\label{eq:radial_coordinate_4DEGBdS_1}
\end{equation}
with
\begin{equation}
\eta=\frac{r_{+}-r_{c}}{r_{+}-r_{--}} \quad (<0).
\label{eq:eta_4DEGBdS_1}
\end{equation}
These definitions move the three singularities $(r_{--},r_{+},r_{c})$ to the points $(0,1,\infty)$; the exterior de Sitter cosmological horizon is now a regular singularity at (spatial) infinity. The next step is to perform an \textit{F-homotopic transformation} $U(x) \mapsto y(x)$ given by
\begin{equation}
U(x)=x^{\frac{\gamma-1}{2}}(x-1)^{\frac{\delta-1}{2}}(x-\eta)^{\frac{\epsilon+1}{2}}y(x),
\label{eq:F-homotopic_4DEGBdS_1}
\end{equation}
where
\begin{eqnarray}
\gamma		& = & 1-\frac{2ir_{--}^{2}\omega}{(r_{c}-r_{--})(r_{+}-r_{--})},\label{eq:gamma_4DEGBdS_1}\\
\delta		& = & 1-\frac{2ir_{+}^{2}\omega}{(r_{c}-r_{+})(r_{+}-r_{--})},\label{eq:delta_4DEGBdS_1}\\
\epsilon	& = & 1-2i\omega.\label{eq:epsilon_4DEGBdS_1}
\end{eqnarray}
Thus, by substituting Eqs.~(\ref{eq:metric_function_4DEGBdS_1})-(\ref{eq:epsilon_4DEGBdS_1}) into Eq.~(\ref{eq:radial_equation_4DEGB}), we get
\begin{equation}
\frac{d^{2} y(x)}{d x^{2}}+\biggl(\frac{\gamma}{x}+\frac{\delta}{x-1}+\frac{\epsilon}{x-\eta}\biggr)\frac{d y(x)}{d x}+\frac{\alpha\beta x-q}{x(x-1)(x-x_{1})}y(x)=0,
\label{eq:radial_final_4DEGBdS_1}
\end{equation}
where
\begin{eqnarray}
\alpha		& = & 1-\frac{2ir_{--}^{2}\omega}{(r_{c}-r_{--})(r_{+}-r_{--})},\label{eq:alpha_4DEGBdS_1}\\
\beta			& = & 1-\frac{2i(r_{c}r_{+}-r_{c}r_{--}+r_{+}r_{--})\omega}{(r_{c}-r_{+})(r_{+}-r_{--})},\label{eq:beta_4DEGBdS_1}\\
q					& = & -\frac{4 r_{--}^3 (2 r_{+}-r_{--}) \omega^2 }{(r_{--}-r_{c}) (r_{+}-r_{--})^3}+\frac{2 i r_{--} (2 r_{+} r_{--}-r_{c} r_{+}-r_{--}^2) \omega }{(r_{--}-r_{c}) (r_{+}-r_{--})^2}+\frac{2 r_{+}-r_{c}-r_{--}-1-3 \lambda}{3 (r_{+}-r_{--})}.\label{eq:q_4DEGBdS_1}
\end{eqnarray}

Equation (\ref{eq:radial_final_4DEGBdS_1}) has the (canonical) form of a general Heun equation, where $y(x) \equiv \mbox{HeunG}(\eta,q;\alpha,\beta,\gamma,\delta;x)$ denotes the general Heun function, which is the solution corresponding to the exponent 0 at $x=0$ and assumes the value 1 there. If $\gamma$ is not a negative integer, then from the Fuchs-Frobenius theory it follows that the $\mbox{HeunG}(\eta ,q;\alpha,\beta,\gamma,\delta;x)$ exists, is analytic in the disk $|x| < 1$, and has a Maclaurin expansion given by
\begin{equation}
\mbox{HeunG}(\eta ,q;\alpha,\beta,\gamma,\delta;x)=\sum_{n=0}^{\infty}c_{n}x^{n},
\label{eq:serie_HeunG_todo_x}
\end{equation}
with
\begin{eqnarray}
-qc_{0}+\eta  \gamma c_{1} & = & 0,\nonumber\\
P_{n}c_{n-1}-(Q_{n}+q)c_{n}+X_{n}c_{n+1} & = & 0 \quad (n \geq 1),
\label{eq:recursion_General_Heun}
\end{eqnarray}
where
\begin{eqnarray}
P_{n} & = & (n-1+\alpha)(n-1+\beta),\nonumber\\
Q_{n} & = & n[(n-1+\gamma)(1+\eta )+\eta \delta+\epsilon],\nonumber\\
X_{n} & = & (n+1)(n+\gamma)\eta .
\label{eq:P_Q_X_recursion_General_Heun}
\end{eqnarray}
Here, the normalization $c_{0}=1$ was adopted by Karl Heun \cite{MathAnn.33.161}. On the other hand, if $\gamma$ is not a positive integer, the solution of Eq.~(\ref{eq:radial_final_4DEGBdS_1}) corresponding to the exponent $1-\gamma$ at $x=0$ is $x^{1-\gamma}\mbox{HeunG}(\eta ,(\eta \delta+\epsilon)(1-\gamma)+q;\alpha+1-\gamma,\beta+1-\gamma,2-\gamma,\delta;x)$.

Therefore, the analytical solution for the radial part of the conformally coupled massless Klein-Gordon equation, in the 4DEGBdS black hole spacetime with $a < 0$, can be written as
\begin{equation}
U_{j}(x)=x^{\frac{\gamma-1}{2}}(x-1)^{\frac{\delta-1}{2}}(x-\eta)^{\frac{\epsilon+1}{2}}[C_{1,j}\ y_{1,j}(x) + C_{2,j}\ y_{2,j}(x)],
\label{eq:analytical_solution_radial_4DEGBdS_1}
\end{equation}
where $C_{1,j}$ and $C_{2,j}$ are constants to be determined, and $j=\{\eta,0,1,\infty\}$ labels the solution at each singular point. Thus, the pair of linearly independent solutions at $x=0$ ($r=r_{--}$) is given by
\begin{eqnarray}
y_{1,0} & = & \mbox{HeunG}(\eta ,q;\alpha,\beta,\gamma,\delta;x),\label{eq:y10_4DEGBdS_1}\\
y_{2,0} & = & x^{1-\gamma}\mbox{HeunG}(\eta ,(\eta \delta+\epsilon)(1-\gamma)+q;\alpha+1-\gamma,\beta+1-\gamma,2-\gamma,\delta;x).\label{eq:y20_4DEGBdS_1}
\end{eqnarray}
Similarly, the pair of linearly independent solutions corresponding to the exponents $0$ and $1-\delta$ at $x=1$ ($r=r_{+}$) is given by
\begin{eqnarray}
y_{1,1} & = & \mbox{HeunG}(1-\eta ,\alpha\beta-q;\alpha,\beta,\delta,\gamma;1-x),\label{eq:y11_4DEGBdS_1}\\
y_{2,1} & = & (1-x)^{1-\delta}\mbox{HeunG}(1-\eta ,((1-\eta )\gamma+\epsilon)(1-\delta)+\alpha\beta-q;\alpha+1-\delta,\beta+1-\delta,2-\delta,\gamma;1-x).\label{eq:y21_4DEGBdS_1}
\end{eqnarray}
The pair of linearly independent solutions corresponding to the exponents $0$ and $1-\epsilon$ at $x=\eta $ ($r=\infty$) is given by
\begin{eqnarray}
y_{1,\eta } & = & \mbox{HeunG}\biggl(\frac{\eta }{\eta -1},\frac{\alpha\beta \eta -q}{\eta -1};\alpha,\beta,\epsilon,\delta;\frac{\eta -x}{\eta -1}\biggl),\label{eq:y1x1_4DEGBdS_1}\\
y_{2,\eta } & = & \biggl(\frac{\eta -x}{\eta -1}\biggl)^{1-\epsilon}\mbox{HeunG}\biggl(\frac{\eta }{\eta -1},\frac{(\eta (\delta+\gamma)-\gamma)(1-\epsilon)}{\eta -1}+\frac{\alpha\beta \eta -q}{\eta -1};\alpha+1-\epsilon,\beta+1-\epsilon,2-\epsilon,\delta;\frac{\eta -x}{\eta -1}\biggl).\nonumber\\\label{eq:y2x1_4DEGBdS_1}
\end{eqnarray}
Finally, the pair of linearly independent solutions corresponding to the exponents $\alpha$ and $\beta$ at $x=\infty$ ($r=r_{c}$) is given by
\begin{eqnarray}
y_{1,\infty} & = & x^{-\alpha}\mbox{HeunG}\biggl(\frac{1}{\eta },\alpha(\beta-\epsilon)+\frac{\alpha}{\eta }(\beta-\delta)-\frac{q}{\eta };\alpha,\alpha-\gamma+1,\alpha-\beta+1,\delta;\frac{1}{x}\biggl),\label{eq:y1i_4DEGBdS_1}\\
y_{2,\infty} & = & x^{-\beta}\mbox{HeunG}\biggl(\frac{1}{\eta },\beta(\alpha-\epsilon)+\frac{\beta}{\eta }(\alpha-\delta)-\frac{q}{\eta };\beta,\beta-\gamma+1,\beta-\alpha+1,\delta;\frac{1}{x}\biggl).\label{eq:y2i_4DEGBdS_1}
\end{eqnarray}

The assumption of a specific asymptotic behavior on the aforementioned analytical solutions near the exterior black hole event horizon ($r \rightarrow r_{+}$) and spatial infinity ($r \rightarrow r_{c}$) can lead to various physical solutions. Next, we will use the VBK approach \cite{AnnPhys.373.28,PhysRevD.104.024035} to derive the characteristic resonance equation and then find the spectrum of resonant frequencies related to quasibound states. In order to compute this quasispectrum, we need to impose two boundary conditions on the radial solution: it should describe an ingoing wave at the exterior black hole event horizon and tend to zero far from the black hole at spatial infinity.
%
%%%%%%%%%%%%%%%%%%%%%%%%%%%%%%%%%%%%%%%%%%%%%%%%%%%%%%%%%%%%%%%%%%%%%%%%%%%%%%%%%%%%%%%%%%%%%% Hawking radiation
%
\subsubsection{Hawking radiation}
In the limit when $r \rightarrow r_{+}$, which implies that $x \rightarrow 1$, the radial solution given by Eq.~(\ref{eq:analytical_solution_radial_4DEGBdS_1}) behaves as
\begin{equation}
\lim_{x \rightarrow 1} U_{1}(x) \sim C_{1,1}\ (x-1)^{\frac{1}{2}(\delta-1)} + C_{2,1}\ (x-1)^{-\frac{1}{2}(\delta-1)},
\label{eq:asymptotic1_4DEGBdS_1}
\end{equation}
where $C_{1,1}$ and $C_{2,1}$ are constants to be determined, in which all remaining constants were included. Then, we can include the time dependence and hence this solution is written as
\begin{equation}
\Psi_{1}(x,t) \sim C_{1,1}\ \Psi_{{\rm in},1} + C_{2,1}\ \Psi_{{\rm out},1}.
\label{eq:Hawking_4DEGBdS_1}
\end{equation}
Here, the ingoing, $\Psi_{{\rm in},1}$, and outgoing, $\Psi_{{\rm out},1}$, scalar wave solutions at the exterior black hole event horizon are given by
\begin{eqnarray}
\Psi_{{\rm in},1}(x>1)	& = & \mbox{e}^{-i \omega t}(x-1)^{-\frac{i\omega}{2|\kappa_{+}|}},\label{eq:Hawking_in_4DEGBdS_1}\\
\Psi_{{\rm out},1}(x>1)	& = & \mbox{e}^{-i \omega t}(x-1)^{+\frac{i\omega}{2|\kappa_{+}|}},\label{eq:Hawking_out_4DEGBdS_1}
\end{eqnarray}
with
\begin{equation}
\frac{1}{2}(\delta-1)=-\frac{i\omega}{2|\kappa_{+}|}.
\label{eq:gamma_Hawking_4DEGBdS_1}
\end{equation}
In addition, the gravitational acceleration, $\kappa_{+}$, on the exterior black hole event horizon is
\begin{equation}
\kappa_{+} \equiv \frac{1}{2r_{+}^{2}} \left.\frac{df(r)}{dr}\right|_{r=r_{+}} = \frac{(r_{+}-r_{c})(r_{+}-r_{--})}{2r_{+}^{2}} \quad (<0).
\label{eq:grav_acc_4DEGBdS_1}
\end{equation}
Thus, by using the analytic continuation described in the VBK approach \cite{AnnPhys.373.28,PhysRevD.104.024035}, we can obtain the relative scattering probability, $\Gamma_{+}$, and the exact spectrum of Hawking-Unruh radiation, $\bar{N}_{\omega}$, given by
\begin{eqnarray}
\Gamma_{+}				& = & \left|\frac{\Psi_{{\rm out},1}(x>1)}{\Psi_{{\rm out},1}(x<1)}\right|^{2}=\mbox{e}^{-\frac{2\pi\omega}{|\kappa_{+}|}},\label{eq:rel_prob_4DEGBdS_1}\\
\bar{N}_{\omega}	& = & \frac{\Gamma_{+}}{1-\Gamma_{+}}=\frac{1}{\mbox{e}^{\omega/k_{B}|T_{+}|}-1},\label{eq:rad_spec_4DEGBdS_1}
\end{eqnarray}
where $k_{B}$ is the Boltzmann constant, and $T_{+}(=|\kappa_{+}|/2\pi k_{B})$ is the Hawking temperature at the exterior black hole event horizon. This spectrum has a thermal character and hence it is analogous to the spectrum of black body radiation.
%
%%%%%%%%%%%%%%%%%%%%%%%%%%%%%%%%%%%%%%%%%%%%%%%%%%%%%%%%%%%%%%%%%%%%%%%%%%%%%%%%%%%%%%%%%%%%%% Quasibound states
%
\subsubsection{Quasibound states}
In order to satisfy purely ingoing boundary conditions, from the asymptotic behavior of the radial solution at the exterior black hole event horizon described by Eq.~(\ref{eq:Hawking_4DEGBdS_1}), we must impose that $C_{2,1}=0$.

Now, let us analyze under which circumstances the radial solution vanishes at spatial infinity. To do this, we have to use the two linearly independent solutions of the general Heun equation at spatial infinity, given by Eqs.~(\ref{eq:y1i_4DEGBdS_1}) and (\ref{eq:y2i_4DEGBdS_1}), and then obtains the following asymptotic behavior for the radial solution
\begin{equation}
\lim_{x \rightarrow \infty} U_{\infty}(x) \sim C_{1,\infty}\ \frac{1}{x^{\sigma}},
\label{eq:radial_infinity_4DEGBdS_1}
\end{equation}
where
\begin{equation}
\sigma=A\omega-1+\alpha,
\label{eq:sigma_4DEGBdS_1}
\end{equation}
with
\begin{equation}
A=\frac{i(r_{c}^2 r_{+}-r_{c}^2 r_{--}+2 r_{c} r_{--}^2-2 r_{+} r_{--}^2)}{(r_{c}-r_{+}) (r_{c}-r_{--}) (r_{+}-r_{--})}.
\label{eq:A_4DEGBdS_1}
\end{equation}
Thus, the radial solution fully satisfies the second boundary condition for QBSs if the sign of the real part of $\sigma$ is such that $\mbox{Re}[\sigma] > 0$; whereas the radial solution diverges at spatial infinity if $\mbox{Re}[\sigma] < 0$. Then, the final asymptotic behavior of the radial solution at spatial infinity will be determined when we know the values of the coefficient $\sigma$, which depends on the frequencies $\omega$, and on the parameter $\alpha$. In order to determine this, we use a polynomial condition for the general Heun functions to match the two asymptotic solutions of the scalar radial equation in their common overlap region. It is known that the general Heun function becomes a (class I) polynomial of degree $n$ if and only if the following two conditions are satisfied \cite{Ronveaux:1995}:
\begin{eqnarray}
\alpha	& = & -n,\label{eq:alpha-condition}\\
c_{n+1}	& = & 0,\label{eq:q-condition}
\end{eqnarray}
where $n=0,1,2,\ldots$. Equation (\ref{eq:alpha-condition}) is the so-called $\alpha$ condition, which provides the expression for the frequencies $\omega_{n}$. On the other hand, the accessory parameter $q$ must be appropriately chosen so that Eqs.~(\ref{eq:recursion_General_Heun}) and (\ref{eq:P_Q_X_recursion_General_Heun}) are consistent, which means that it is also necessary for the accessory parameter $q$ to be an eigenvalue of the general Heun equation, calculated via Eq.~(\ref{eq:q-condition}). Furthermore, the accessory parameter $q$ contains the separation constant $\lambda$, which indicates that we could obtain the eigenvalues of the separation constant $\lambda_{n}(\omega_{n})$, corresponding to the appropriate eigenvalues of the frequencies $\omega_{n}$, from the polynomial solution for the radial equation and then use it to show the (regular) angular behavior of massless scalar QBSs in the background under consideration (for details, see Ref. \cite{PhysRevD.105.045015}).

In the present case, by imposing the condition given by Eq.~(\ref{eq:alpha-condition}), with the parameter $\alpha$ given by Eq.~(\ref{eq:alpha_4DEGBdS_1}), we obtain a first order equation for $\omega$, whose solution is the exact spectrum of QBSs:
\begin{equation}
\omega_{n}=-\frac{i(n+1)(r_{c}-r_{--})(r_{+}-r_{--})}{2r_{--}^{2}},
\label{eq:omega_4DEGBdS_1}
\end{equation}
where $n$ can be called the principal quantum number. As a result, the quasispectrum of frequency eigenvalues of the conformally coupled massless Klein-Gordon equation (\ref{eq:conformal_massless_KGE}) is discrete, that is, the frequency quasispectrum consists of an infinite number of discrete frequency levels corresponding to quasibound states. Furthermore, note that it is a function of the horizons $r_{j}$, and therefore the QBSs are also function of continuous variables, such as the cosmological constant, GB coupling constant, and cosmological radius.

\begin{figure}[p]
\centering
\includegraphics[width=1\columnwidth]{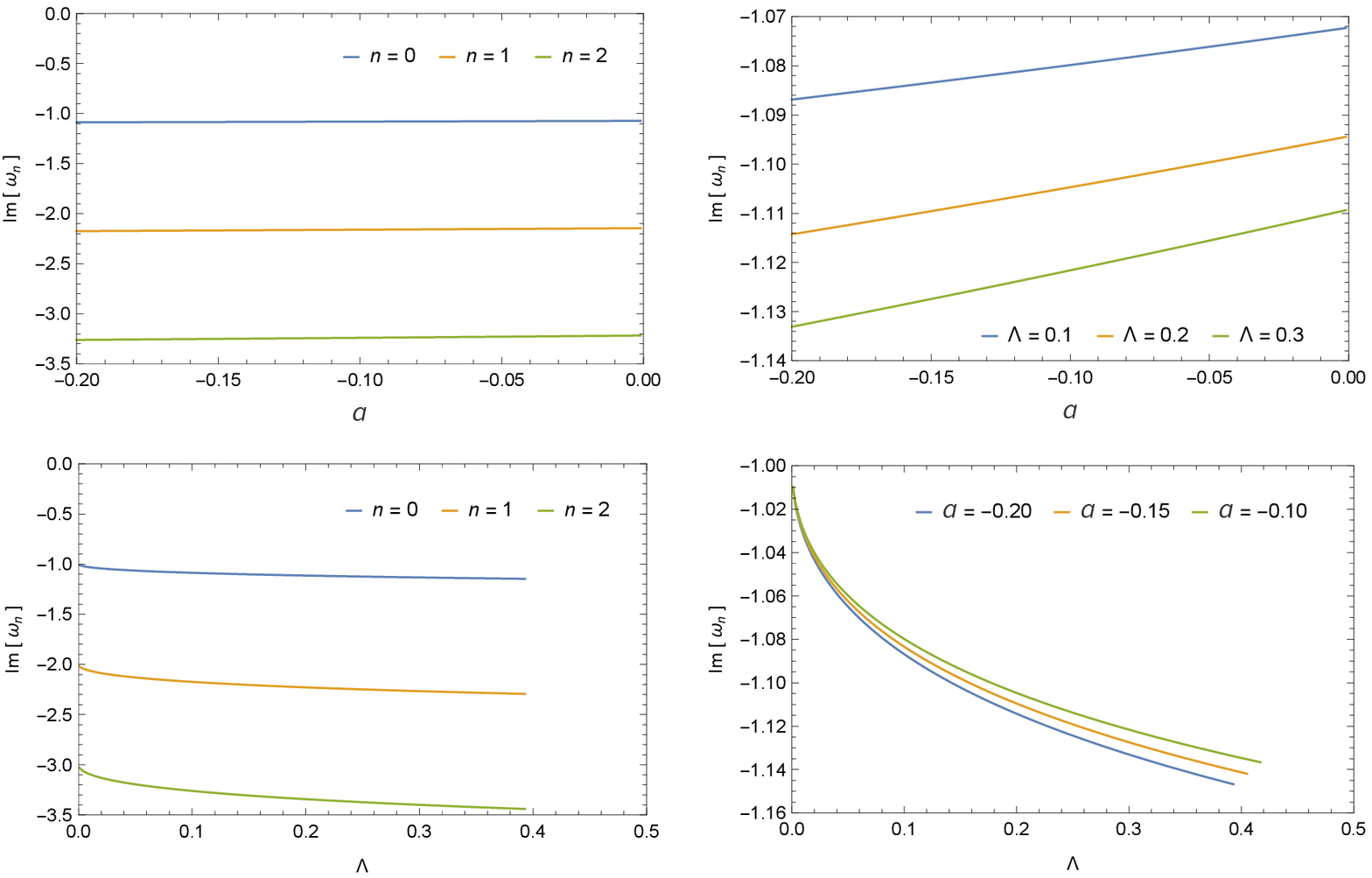}
\caption{(dS) Case 1: $a < 0$. Top panel: massless scalar QBSs as a function of the GB coupling constant for $\Lambda=0.1$ (left) and $n=0$ (right); $M=1/2$. Bottom panel: massless scalar QBSs as a function of the cosmological constant for $a=-0.20$ (left) and $n=0$ (right); $M=1/2$.}
\label{fig:Fig2_4DEGBdS}
\end{figure}

In Table \ref{tab:I_4DEGBdS} we present the QBSs $\omega_{n}$ as functions of the GB coupling constant $a$, for some values of the cosmological constant $\Lambda$. In addition, we also present the behavior of the QBSs $\omega_{n}$ in Fig.~\ref{fig:Fig2_4DEGBdS}. We can conclude that the spectrum given by Eq.~(\ref{eq:omega_4DEGBdS_1}) is physically admissible, since the real part of $\sigma$ is positive, and therefore it represents QBS frequencies for conformally coupled massless scalars in the 4DEGBdS black hole spacetime with $a < 0$. Furthermore, we can see that the decay is overdamped, since this quasispectrum is purely imaginary.

\begin{table}[t]
	\caption{(dS) Cases 1 and 2. Massless scalar fundamental QBSs $\omega_{n}$ and their corresponding values of the real part of $\sigma_{n}=A\omega_{n}-1-n$, and the angular eigenvalue $\lambda_{0;1}$ and their corresponding general degree $\nu_{0;1}$, for various values of the GB coupling constant in the stability sector; $n=0$, $M=1/2$.}
	\label{tab:I_4DEGBdS}
	\begin{tabular}{c|c|c|c|c}
		\hline\noalign{\smallskip}
		$a$			& $\omega_{0}$	& $\sigma_{0}$	& $\lambda_{0;1}$	& $\nu_{0;1}$			\\
		\noalign{\smallskip}\hline\noalign{\smallskip}
		\multicolumn{5}{c}{$\Lambda=0.1$} 			\\
		\noalign{\smallskip}\hline\noalign{\smallskip}
		$-0.20$	& $-1.08687i$		& $0.640242$		& $6.57866$				& $2.11317$				\\
		$-0.15$	& $-1.08342i$		& $0.634229$		& $6.54704$				& $2.10711$				\\
		$-0.10$	& $-1.07985i$		& $0.628086$		& $6.51415$				& $2.10080$				\\
		$-0.05$	& $-1.07612i$		& $0.621792$		& $6.47982$				& $2.09419$				\\
		$0.05$	& $-21615.1i$		& $2532.930$		& $-1119.55$			& $-0.5-33.4559i$	\\
		$0.10$	& $-4815.11i$		& $562.8010$		& $-518.670$			& $-0.5-22.7688i$	\\
		$0.15$	& $-1881.80i$		& $219.3870$		& $-317.687$			& $-0.5-17.8168i$	\\
		$0.20$	& $-915.112i$		& $106.4160$		& $-216.565$			& $-0.5-14.7076i$	\\
		\noalign{\smallskip}\hline\noalign{\smallskip}
		\multicolumn{5}{c}{$\Lambda=0.2$} 			\\
		\noalign{\smallskip}\hline\noalign{\smallskip}
		$-0.20$	& $-1.11428i$		& $0.760966$		& $4.90808$				& $1.77114$				\\
		$-0.15$	& $-1.10957i$		& $0.747361$		& $4.87752$				& $1.76440$				\\
		$-0.10$	& $-1.10470i$		& $0.733818$		& $4.84573$				& $1.75737$				\\
		$-0.05$	& $-1.09963i$		& $0.720291$		& $4.81255$				& $1.75001$				\\
		$0.05$	& $-10807.6i$		& $2152.410$		& $-559.940$			& $-0.5-23.6578i$	\\
		$0.10$	& $-2407.64i$		& $475.4460$		& $-259.506$			& $-0.5-16.1014i$	\\
		$0.15$	& $-941.027i$		& $184.2920$		& $-159.021$			& $-0.5-12.6004i$	\\
		$0.20$	& $-457.733i$		& $88.91780$		& $-108.470$			& $-0.5-10.4029i$	\\
		\noalign{\smallskip}\hline\noalign{\smallskip}
		\multicolumn{5}{c}{$\Lambda=0.3$} 			\\
		\noalign{\smallskip}\hline\noalign{\smallskip}
		$-0.20$	& $-1.13308i$		& $0.997463$		& $4.15273$				& $1.59827$				\\
		$-0.15$	& $-1.12744i$		& $0.957611$		& $4.12281$				& $1.59113$				\\
		$-0.10$	& $-1.12161i$		& $0.920943$		& $4.09171$				& $1.58368$				\\
		$-0.05$	& $-1.11555i$		& $0.886856$		& $4.05926$				& $1.57587$				\\
		$0.05$	& $-7205.07i$		& $2234.500$		& $-373.405$			& $-0.5-19.3172i$	\\
		$0.10$	& $-1605.15i$		& $486.2310$		& $-173.118$			& $-0.5-13.1479i$	\\
		$0.15$	& $-627.436i$		& $185.8920$		& $-106.132$			& $-0.5-10.2899i$	\\
		$0.20$	& $-305.274i$		& $88.56510$		& $-72.4383$			& $-0.5-8.49637i$	\\
		\noalign{\smallskip}\hline
	\end{tabular}
\end{table}

%
%%%%%%%%%%%%%%%%%%%%%%%%%%%%%%%%%%%%%%%%%%%%%%%%%%%%%%%%%%%%%%%%%%%%%%%%%%%%%%%%%%%%%%%%%%%%%% Radial wave eigenfunctions
%
\subsubsection{Radial wave eigenfunctions}
The radial wave eigenfunctions, which are related to QBSs of conformally coupled massless scalars propagating in the 4DEGBdS black hole spacetime with $a < 0$, can be obtained by using the condition given by Eq.~(\ref{eq:q-condition}), and imposing the condition given by Eq.~(\ref{eq:alpha-condition}), as it is described in the VBK approach \cite{AnnPhys.373.28,PhysRevD.104.024035}.

As we explained before, these polynomial radial eigenfunctions are related to the appropriate determination of the eigenvalue $q$. Since it is calculated via Eq.~(\ref{eq:q-condition}), we index its solutions by a parameter $s(=1,2,3,\ldots)$, which can be conveniently denoted by $q_{n;s}$. Thus, the corresponding general Heun polynomials are now denoted as $\mbox{HeunGp}_{n;s}(x)$. Therefore, the QBS radial wave eigenfunctions for conformally coupled massless scalars propagating in a 4DEGBdS black hole spacetime with $a < 0$ are given by
\begin{equation}
U_{n;s}(x)=C_{n;s}\ x^{\frac{\gamma-1}{2}}(x-1)^{\frac{\delta-1}{2}}(x-\eta)^{\frac{\epsilon+1}{2}}\ \mbox{HeunGp}_{n;s}(\eta,q_{n,s};-n,\beta,\gamma,\delta;x),
\label{eq:radial_eigenfunctions_4DEGBdS_1}
\end{equation}
where $C_{n;s}$ is a constant to be determined. 

Next, we calculate the general Heun polynomials related to the fundamental and first excited modes. The general Heun polynomial for the fundamental mode $n=0$ is given by
\begin{equation}
\mbox{HeunGp}_{0;1}(x)=1,
\label{eq:Hp_0,0}
\end{equation}
where the eigenvalue $q_{0;1}$ must obey
\begin{equation}
c_{1}=\frac{q}{\eta\gamma}=0,
\label{eq:c_1}
\end{equation}
whose unique solution ($s=1$) is
\begin{equation}
q_{0;1}=0.
\label{eq:q_0,1}
\end{equation}
On the other hand, the general Heun polynomials for the first excited mode $n=1$ are given by
\begin{equation}
\mbox{HeunGp}_{1;s}(x)=c_{0}+c_{1}x=1+\frac{q_{1;s}}{\eta\gamma}x,
\label{eq:Hp_1,s}
\end{equation}
where the eigenvalues $q_{1;s}$ must obey
\begin{equation}
c_{2}=\frac{[\gamma(1+\eta)+\eta\delta+\epsilon+q]q-\eta\alpha\beta\gamma}{2\eta^{2}\gamma(1+\gamma)}=0,
\label{eq:c_2}
\end{equation}
whose two solutions ($s=1,2$) are
\begin{equation}
q_{1;1}=\frac{-[\gamma(1+\eta)+\eta\delta+\epsilon] - \sqrt{\Delta}}{2},
\label{eq:q_1,1}
\end{equation}
\begin{equation}
q_{1;2}=\frac{-[\gamma(1+\eta)+\eta\delta+\epsilon] + \sqrt{\Delta}}{2},
\label{eq:q_1,2}
\end{equation}
where $\Delta=[\gamma(1+\eta)+\eta\delta+\epsilon]^{2}+4\eta\alpha\beta\gamma$. It is worth noticing that the polynomial radial eigenfunctions for the first excited mode $n=1$ are degenerate, since there exist two solutions for the eigenvalue $q$. However, we can choose one of these solutions by analyzing their behaviors. Indeed, the first solution ($s=1$), and its corresponding eigenvalue ($q_{1;1}$) as well, is not suitable for our description and therefore it does not describe quasibound states, that is, it does not fully satisfy the two boundary conditions, since it has a false (nonphysical) singular point in the region between the exterior black hole event horizon and spatial infinity.

\begin{figure}[p]
\centering
\includegraphics[width=1\columnwidth]{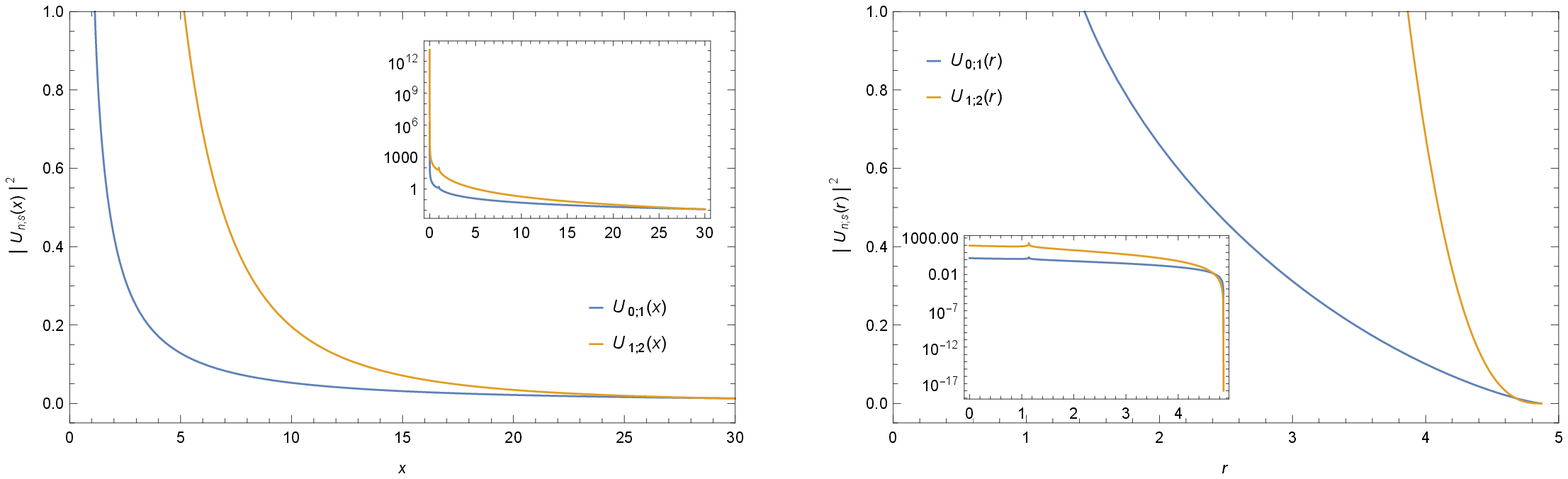}
\caption{(dS) Case 1: $a < 0$. The first two squared QBS radial wave eigenfunctions as a function of the new radial coordinate $x$ (left) and radial coordinate $r$ (right); $a=-0.20$, $\Lambda=0.1$, $M=1/2$. The units are in multiples of $C_{n;s}$.}
\label{fig:Fig3_4DEGBdS}
\end{figure}

In Fig.~\ref{fig:Fig3_4DEGBdS}, we plot the first two squared QBS radial wave eigenfunctions. We observe that the radial solution tends to zero at spatial infinity ($x \rightarrow \infty$ or $r_{c}=4.86958$, for $a=-0.20$, and $\Lambda=0.1$) and diverges at the exterior black hole event horizon, which represents QBSs. Note that the radial wave eigenfunctions reach a maximum value at the exterior black hole event horizon ($x=1$ or $r_{+}=1.13694$, for $a=-0.20$, and $\Lambda=0.1$), and then cross this surface, as shown in the log-scale plots.

%
%%%%%%%%%%%%%%%%%%%%%%%%%%%%%%%%%%%%%%%%%%%%%%%%%%%%%%%%%%%%%%%%%%%%%%%%%%%%%%%%%%%%%%%%%%%%%% Angular wave eigenfunctions
%
\subsubsection{Angular wave eigenfunctions}
The radial equation (\ref{eq:radial_equation_4DEGB}) and the angular equation (\ref{eq:angular_equation_4DEGB}) are coupled by a discrete set of angular eigenvalues $\lambda$, which are solutions due to the regularity requirements imposed to the angular functions at the two boundaries $\theta=0$ and $\theta=\pi$.

For the QBSs, we can obtain an expression for the angular eigenvalues $\lambda_{n;s}(\omega_{n})$ from the polynomial equation (\ref{eq:q-condition}) that determined the accessory parameter $q_{n;s}$. Let us focus on the fundamental mode $n=0$. Thus, from Eqs.~(\ref{eq:q_4DEGBdS_1}) and (\ref{eq:q_0,1}), we obtain the following expression for the angular eigenvalues with $a < 0$:
\begin{equation}
\lambda_{0;1}=-\frac{4 r_{--}^3 (r_{--}-2 r_{+}) \omega_{0}^2 }{(r_{c}-r_{--}) (r_{--}-r_{+})^2}-\frac{2 i r_{--} (r_{c} r_{+}-2 r_{+} r_{--}+r_{--}^2) \omega_{0} }{(r_{c}-r_{--}) (r_{--}-r_{+})}+\frac{1}{3} (2 r_{+}-r_{c}-r_{--}-1).
\label{eq:lambda_01_4DEGBdS_1}
\end{equation}
In this case, the angular eigenvalues $\lambda_{0;1}$ are real and positive, and hence the general degree $\nu_{0;1}$ will be positive real number, which can be numerically evaluated from Eq.~(\ref{eq:lambda_01_4DEGBdS_1}).

\begin{figure}[p]
\centering
\includegraphics[scale=1]{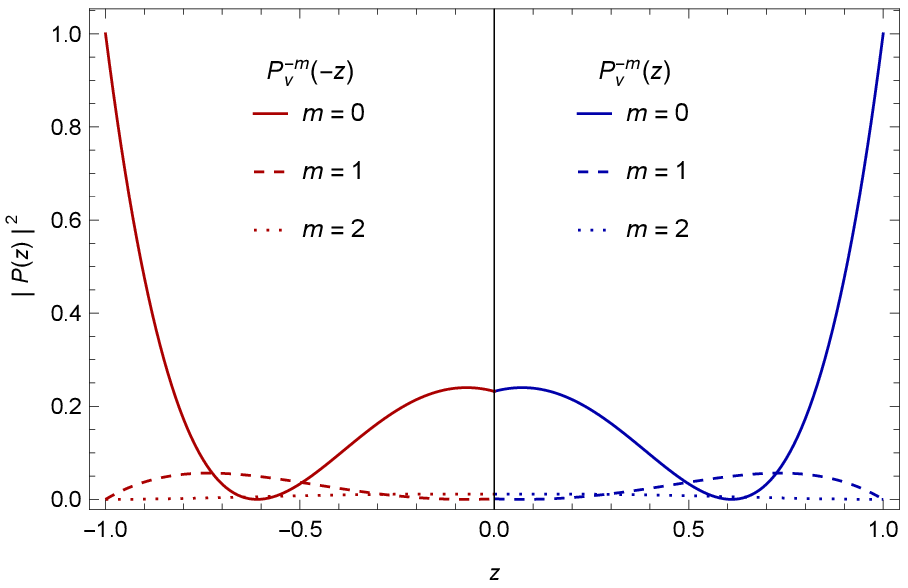}
\caption{(dS) Case 1: $a < 0$. The first three squared QBS angular wave eigenfunctions as a function of the new polar coordinate $z$; $a=-0.20$, $\Lambda=0.1$, $n=0$, $M=1/2$.}
\label{fig:Fig4_4DEGBdS}
\end{figure}

In Table \ref{tab:I_4DEGBdS}, we present the angular eigenvalues $\lambda_{0;1}$, as well as the corresponding general degree $\nu_{0;1}$, as functions of the GB coupling constant $a$, for some values of the cosmological constant $\Lambda$. In addition, we also present the behavior of the QBS angular wave eigenfunctions $P(\theta)$ in Fig.~\ref{fig:Fig4_4DEGBdS}, as functions of the new polar coordinate $z=\cos\theta$, for some values of the magnetic quantum number $m$. It is worth emphasizing that the two numerically satisfactory solutions of the associated Legendre equation of general degree are given in terms of the Ferrers function of the first kind $P_{\nu}^{-m}(-z)$ and $P_{\nu}^{-m}(z)$ in the interval $-1 < x < 1$. Note that these angular solutions are regular at the two boundaries $\theta=\pi$ ($z=-1$) and $\theta=0$ ($z=1$).

%
%%%%%%%%%%%%%%%%%%%%%%%%%%%%%%%%%%%%%%%%%%%%%%%%%%%%%%%%%%%%%%%%%%%%%%%%%%%%%%%%%%%%%%%%%%%%%% (dS) Case 2: a > 0
%
\subsection{(dS) Case 2: \texorpdfstring{$a > 0$}{a > 0}}
In this case, we adopt the following parametrization for the metric function,
\begin{equation}
f(r)=\frac{1}{r^{2}}(r-r_{c})(r-r_{+})(r-r_{-})(r-r_{--}),
\label{eq:metric_function_4DEGBdS_2}
\end{equation}
where $r_{c}$ is the (positive) ``exterior'' de Sitter cosmological horizon, $r_{+}$ is the (positive) ``exterior'' black hole event horizon, $r_{-}$ is the (positive) ``interior'' black hole event horizon, and $r_{--}$ is the (negative) ``interior'' de Sitter cosmological horizon. Then, with the metric function given by Eq.~(\ref{eq:metric_function_4DEGBdS_2}), it is easy to see that Eq.~(\ref{eq:radial_equation_4DEGB}) is also a (general) Heun-type equation with four finite regular singularities (associated to the four event horizons) and one regular singularity at (spatial) infinity.

Therefore, by following straightforwardly the steps described in the previous section, the analytical solution for the radial part of the conformally coupled massless Klein-Gordon equation, in the 4DEGBdS black hole spacetime with $a > 0$, can be written as
\begin{equation}
U_{j}(x)=x^{\frac{\gamma-1}{2}}(x-1)^{\frac{\delta-1}{2}}(x-x_{4})^{\frac{\epsilon-1}{2}}(x-\eta)[C_{1,j}\ y_{1,j}(x) + C_{2,j}\ y_{2,j}(x)],
\label{eq:analytical_solution_radial_4DEGBdS_2}
\end{equation}
where $C_{1,j}$ and $C_{2,j}$ are constants to be determined, and $j=\{x_{4},0,1,\infty\}$ labels the solution at each singular point. Thus, the pair of linearly independent solutions corresponding to the exponents $0$ and $1-\epsilon$ at $x=x_{4} $ ($r=r_{--}$) is given by
\begin{eqnarray}
y_{1,x_{4} } & = & \mbox{HeunG}\biggl(\frac{x_{4} }{x_{4} -1},\frac{\alpha\beta x_{4} -q}{x_{4} -1};\alpha,\beta,\epsilon,\delta;\frac{x_{4} -x}{x_{4} -1}\biggl),\label{eq:y1x1_4DEGBdS_2}\\
y_{2,x_{4} } & = & \biggl(\frac{x_{4} -x}{x_{4} -1}\biggl)^{1-\epsilon}\mbox{HeunG}\biggl(\frac{x_{4} }{x_{4} -1},\frac{(x_{4} (\delta+\gamma)-\gamma)(1-\epsilon)}{x_{4} -1}+\frac{\alpha\beta x_{4} -q}{x_{4} -1};\alpha+1-\epsilon,\beta+1-\epsilon,2-\epsilon,\delta;\frac{x_{4} -x}{x_{4} -1}\biggl).\nonumber\\\label{eq:y2x1_4DEGBdS_2}
\end{eqnarray}
Similarly, the pair of linearly independent solutions at $x=0$ ($r=r_{-}$) is given by
\begin{eqnarray}
y_{1,0} & = & \mbox{HeunG}(x_{4} ,q;\alpha,\beta,\gamma,\delta;x),\label{eq:y10_4DEGBdS_2}\\
y_{2,0} & = & x^{1-\gamma}\mbox{HeunG}(x_{4} ,(x_{4} \delta+\epsilon)(1-\gamma)+q;\alpha+1-\gamma,\beta+1-\gamma,2-\gamma,\delta;x).\label{eq:y20_4DEGBdS_2}
\end{eqnarray}
Furthermore, the pair of linearly independent solutions corresponding to the exponents $0$ and $1-\delta$ at $x=1$ ($r=r_{+}$) is given by
\begin{eqnarray}
y_{1,1} & = & \mbox{HeunG}(1-x_{4} ,\alpha\beta-q;\alpha,\beta,\delta,\gamma;1-x),\label{eq:y11_4DEGBdS_2}\\
y_{2,1} & = & (1-x)^{1-\delta}\mbox{HeunG}(1-x_{4} ,((1-x_{4} )\gamma+\epsilon)(1-\delta)+\alpha\beta-q;\alpha+1-\delta,\beta+1-\delta,2-\delta,\gamma;1-x).\label{eq:y21_4DEGBdS_2}
\end{eqnarray}
Finally, the pair of linearly independent solutions corresponding to the exponents $\alpha$ and $\beta$ at $x=\infty$ ($r=r_{c}$) is given by
\begin{eqnarray}
y_{1,\infty} & = & x^{-\alpha}\mbox{HeunG}\biggl(\frac{1}{x_{4} },\alpha(\beta-\epsilon)+\frac{\alpha}{x_{4} }(\beta-\delta)-\frac{q}{x_{4} };\alpha,\alpha-\gamma+1,\alpha-\beta+1,\delta;\frac{1}{x}\biggl),\label{eq:y1i_4DEGBdS_2}\\
y_{2,\infty} & = & x^{-\beta}\mbox{HeunG}\biggl(\frac{1}{x_{4} },\beta(\alpha-\epsilon)+\frac{\beta}{x_{4} }(\alpha-\delta)-\frac{q}{x_{4} };\beta,\beta-\gamma+1,\beta-\alpha+1,\delta;\frac{1}{x}\biggl).\label{eq:y2i_4DEGBdS_2}
\end{eqnarray}
Here, the new radial coordinate, $x$, is defined as
\begin{equation}
x=\frac{r_{-}-r}{r_{c}-r}\eta,
\label{eq:radial_coordinate_4DEGBdS_2}
\end{equation}
with
\begin{equation}
\eta=\frac{r_{+}-r_{c}}{r_{+}-r_{-}} \quad (<0).
\label{eq:eta_4DEGBdS_2}
\end{equation}
In addition, we set a new parameter, $x_{4}$, which is associated with the four finite regular singularities through the relation
\begin{equation}
x_{4}=\frac{r_{-}-r_{--}}{r_{c}-r_{--}}\eta \quad (<0).
\label{eq:x_4_4DEGBdS_2}
\end{equation}
The parameters $\alpha$, $\beta$, $\gamma$, $\delta$, $\epsilon$, and $q$ are given by
\begin{eqnarray}
\alpha		& = & 1-\frac{2ir_{-}^{2}\omega}{(r_{-}-r_{c})(r_{-}-r_{+})(r_{-}-r_{--})},\label{eq:alpha_4DEGBdS_2}\\
\beta			& = & 1-\frac{2i\{r_{+} r_{-} r_{--}-r_{c} [r_{+} (r_{-}-r_{--})+r_{-} r_{--}]\}\omega}{(r_{c}-r_{+}) (r_{c}-r_{--}) (r_{-}-r_{+}) (r_{-}-r_{--})},\label{eq:beta_4DEGBdS_2}\\
\gamma		& = & 1-\frac{2ir_{-}^{2}\omega}{(r_{-}-r_{c})(r_{-}-r_{+})(r_{-}-r_{--})},\label{eq:gamma_4DEGBdS_2}\\
\delta		& = & 1+\frac{2ir_{+}^{2}\omega}{(r_{+}-r_{c})(r_{+}-r_{-})(r_{+}-r_{--})},\label{eq:delta_4DEGBdS_2}\\
\epsilon	& = & 1+\frac{2ir_{--}^{2}\omega}{(r_{--}-r_{c})(r_{--}-r_{+})(r_{--}-r_{-})},\label{eq:epsilon_4DEGBdS_2}\\
q					& = & -\frac{4 r_{-}^3 (r_{+} r_{-}-2 r_{+} r_{--}+r_{-} r_{--}) \omega^2 }{(r_{-}-r_{c}) (r_{c}-r_{--}) (r_{+}-r_{-})^3 (r_{-}-r_{--})^2}+\frac{2 i r_{-} (r_{c} r_{+} r_{--}-r_{c} r_{-}^2+r_{+} r_{-}^2-2 r_{+} r_{-} r_{--}+r_{-}^2 r_{--}) \omega }{(r_{-}-r_{c}) (r_{c}-r_{--}) (r_{+}-r_{-})^2 (r_{-}-r_{--})}\nonumber\\
					&		&	+\frac{r_{c} r_{+}-2 r_{c} r_{-}+r_{c} r_{--}+r_{+} r_{-}-2 r_{+} r_{--}+r_{-} r_{--}-1-3 \lambda}{3 (r_{c}-r_{--}) (r_{+}-r_{-})}.\label{eq:q_4DEGBdS_2}
\end{eqnarray}
%
%%%%%%%%%%%%%%%%%%%%%%%%%%%%%%%%%%%%%%%%%%%%%%%%%%%%%%%%%%%%%%%%%%%%%%%%%%%%%%%%%%%%%%%%%%%%%% Hawking radiation
%
\subsubsection{Hawking radiation}
In the limit when $r \rightarrow r_{+}$, which implies that $x \rightarrow 1$, the radial solution given by Eq.~(\ref{eq:analytical_solution_radial_4DEGBdS_2}) behaves as
\begin{equation}
\Psi_{1}(x,t) \sim C_{1,1}\ \Psi_{{\rm in},1} + C_{2,1}\ \Psi_{{\rm out},1},
\label{eq:Hawking_4DEGBdS_2}
\end{equation}
where
\begin{eqnarray}
\Psi_{{\rm in},1}(x>1)	& = & \mbox{e}^{-i \omega t}(x-1)^{-\frac{i\omega}{2|\kappa_{+}|}},\label{eq:Hawking_in_4DEGBdS_2}\\
\Psi_{{\rm out},1}(x>1)	& = & \mbox{e}^{-i \omega t}(x-1)^{+\frac{i\omega}{2|\kappa_{+}|}},\label{eq:Hawking_out_4DEGBdS_2}
\end{eqnarray}
with
\begin{equation}
\frac{1}{2}(\delta-1)=-\frac{i\omega}{2|\kappa_{+}|},
\label{eq:gamma_Hawking_4DEGBdS_2}
\end{equation}
and
\begin{equation}
\kappa_{+} \equiv \frac{1}{2r_{+}^{2}} \left.\frac{df(r)}{dr}\right|_{r=r_{+}} = \frac{(r_{+}-r_{c})(r_{+}-r_{-})(r_{+}-r_{--})}{2r_{+}^{2}} \quad (<0).
\label{eq:grav_acc_4DEGBdS_2}
\end{equation}
Therefore, these wave solutions give a thermal spectrum analogous to the spectrum of black body radiation.
%
%%%%%%%%%%%%%%%%%%%%%%%%%%%%%%%%%%%%%%%%%%%%%%%%%%%%%%%%%%%%%%%%%%%%%%%%%%%%%%%%%%%%%%%%%%%%%% Quasibound states
%
\subsubsection{Quasibound states}
In order to satisfy purely ingoing boundary conditions, from the asymptotic behavior of the radial solution at the exterior black hole event horizon described by Eq.~(\ref{eq:Hawking_4DEGBdS_2}), we must impose that $C_{2,1}=0$.

On the other hand, in the limit when $r \rightarrow r_{c}$, which implies that $x \rightarrow \infty$, the radial solution given by Eq.~(\ref{eq:analytical_solution_radial_4DEGBdS_2}) behaves as
\begin{equation}
\lim_{x \rightarrow \infty} U_{\infty}(x) \sim C_{1,\infty}\ \frac{1}{x^{\sigma}},
\label{eq:radial_infinity_4DEGBdS_2}
\end{equation}
where
\begin{equation}
\sigma=A\omega-1+\alpha,
\label{eq:sigma_4DEGBdS_2}
\end{equation}
with
\begin{equation}
A=\frac{i(r_{c}^2 r_{+} r_{-}-r_{c}^2 r_{+} r_{--}+r_{c}^2 r_{-}^2+r_{c}^2 r_{-} r_{--}-2 r_{c} r_{+} r_{-}^2-2 r_{c} r_{-}^2 r_{--}+2 r_{+} r_{-}^2 r_{--})}{(r_{c}-r_{+}) (r_{c}-r_{-}) (r_{c}-r_{--}) (r_{+}-r_{-}) (r_{-}-r_{--})}.
\label{eq:A_4DEGBdS_2}
\end{equation}
In the present case, by imposing the condition given by Eq.~(\ref{eq:alpha-condition}), with the parameter $\alpha$ given by Eq.~(\ref{eq:alpha_4DEGBdS_2}), we obtain a first order equation for $\omega$, whose solution is the exact spectrum of QBSs:
\begin{equation}
\omega_{n}=-\frac{i(n+1)(r_{-}-r_{c})(r_{-}-r_{+})(r_{-}-r_{--})}{2r_{-}^{2}}.
\label{eq:omega_4DEGBdS_2}
\end{equation}

\begin{figure}[p]
\centering
\includegraphics[width=1\columnwidth]{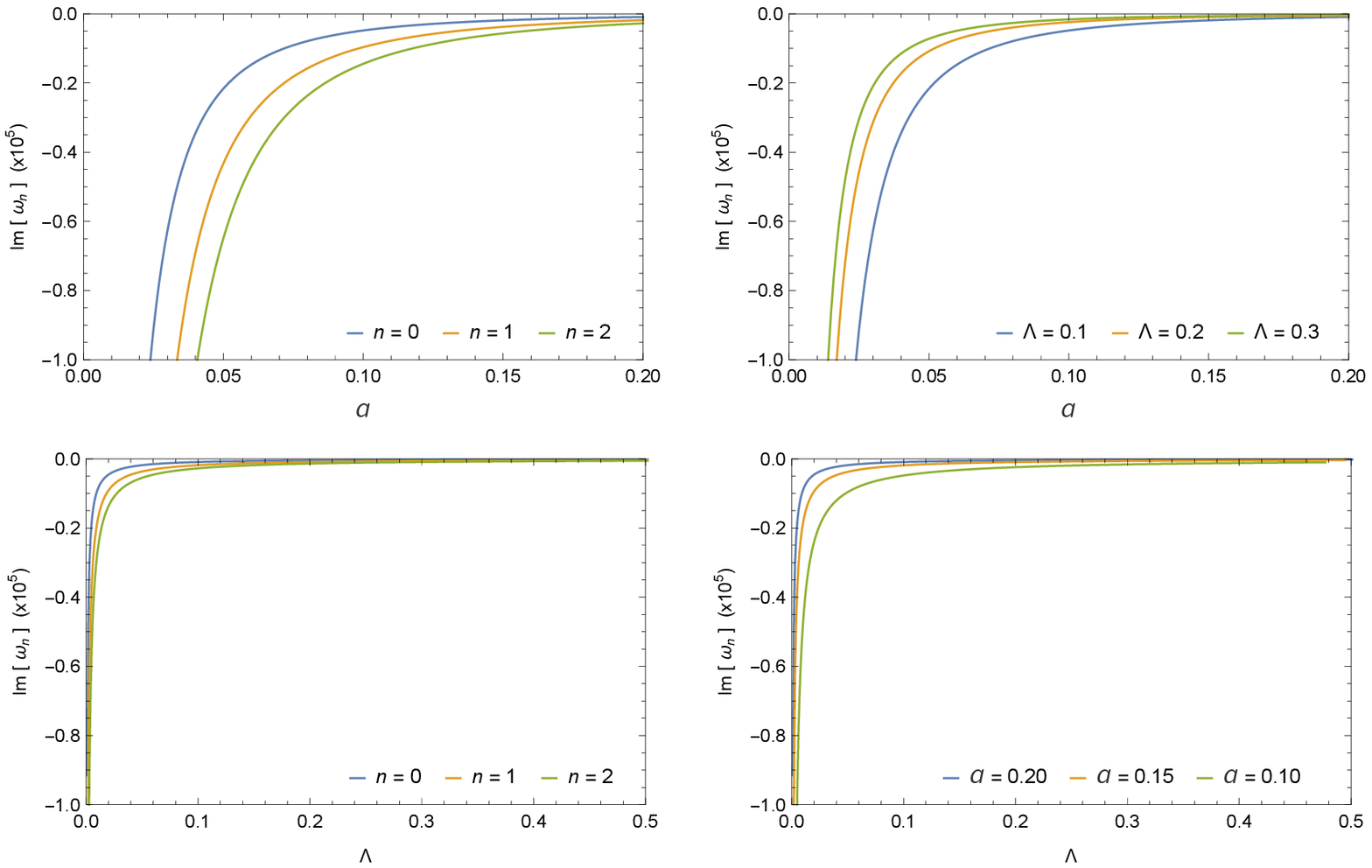}
\caption{(dS) Case 2: $a > 0$. Top panel: massless scalar QBSs as a function of the GB coupling constant for $\Lambda=0.1$ (left) and $n=0$ (right); $M=1/2$. Bottom panel: massless scalar QBSs as a function of the cosmological constant for $a=0.20$ (left) and $n=0$ (right); $M=1/2$.}
\label{fig:Fig5_4DEGBdS}
\end{figure}

In Table \ref{tab:I_4DEGBdS} we present the QBSs $\omega_{n}$ as functions of the GB coupling constant, $a$, for some values of the cosmological constant $\Lambda$. In addition, we also present the behavior of the QBSs $\omega_{n}$ in Fig.~\ref{fig:Fig5_4DEGBdS}. We can conclude that the spectrum given by Eq.~(\ref{eq:omega_4DEGBdS_2}) is physically admissible, since the real part of $\sigma$ is positive, and therefore it represents QBS frequencies for conformally coupled massless scalars in the 4DEGBdS black hole spacetime with $a > 0$. Furthermore, we can see that the decay is overdamped, since this quasispectrum is purely imaginary.

%
%%%%%%%%%%%%%%%%%%%%%%%%%%%%%%%%%%%%%%%%%%%%%%%%%%%%%%%%%%%%%%%%%%%%%%%%%%%%%%%%%%%%%%%%%%%%%% Radial wave eigenfunctions
%
\subsubsection{Radial wave eigenfunctions}
In the present case, the QBS radial wave eigenfunctions for conformally coupled massless scalars propagating in a 4DEGBdS black hole spacetime with $a > 0$ are given by
\begin{equation}
U_{n;s}(x)=C_{n;s}\ x^{\frac{\gamma-1}{2}}(x-1)^{\frac{\delta-1}{2}}(x-x_{4})^{\frac{\epsilon-1}{2}}(x-\eta)\ \mbox{HeunGp}_{n;s}(x_{4},q_{n,s};-n,\beta,\gamma,\delta;x).
\label{eq:radial_eigenfunctions_4DEGBdS_2}
\end{equation}

\begin{figure}[p]
\centering
\includegraphics[width=1\columnwidth]{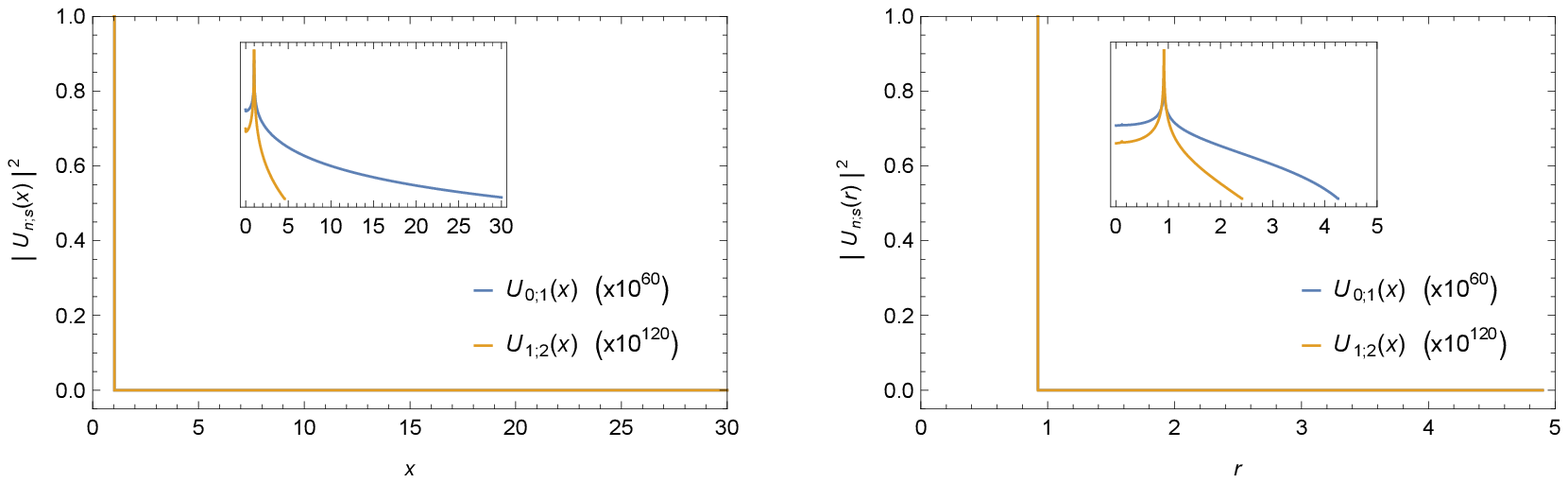}
\caption{(dS) Case 2: $a > 0$. The first two squared QBS radial wave eigenfunctions as a function of the new radial coordinate $x$ (left) and radial coordinate $r$ (right); $a=0.20$, $\Lambda=0.1$, $M=1/2$. The units are in multiples of $C_{n;s}$.}
\label{fig:Fig6_4DEGBdS}
\end{figure}

In Fig.~\ref{fig:Fig6_4DEGBdS}, we plot the first two squared QBS radial wave eigenfunctions. We observe that the radial solution tends to zero at spatial infinity ($x \rightarrow \infty$ or $r_{c}=4.899130$, for $a=0.20$, and $\Lambda=0.1$) and diverges at the exterior black hole event horizon, which represents QBSs. Note that the radial wave eigenfunctions reach a maximum value at the exterior black hole event horizon ($x=1$ or $r_{+}=0.916563$, for $a=0.20$, and $\Lambda=0.1$), and then cross this surface, as shown in the log-scale plots.

%
%%%%%%%%%%%%%%%%%%%%%%%%%%%%%%%%%%%%%%%%%%%%%%%%%%%%%%%%%%%%%%%%%%%%%%%%%%%%%%%%%%%%%%%%%%%%%% Angular wave eigenfunctions
%
\subsubsection{Angular wave eigenfunctions}
In the present case, from Eqs.~(\ref{eq:q_4DEGBdS_2}) and (\ref{eq:q_0,1}), we obtain the following expression for the fundamental mode angular eigenvalues with $a > 0$:
\begin{eqnarray}
\lambda_{0;1} & = & -\frac{4 r_{-}^3 (r_{+} r_{-}-2 r_{+} r_{--}+r_{-} r_{--}) \omega_{0}^2}{(r_{-}-r_{c}) (r_{+}-r_{-})^2 (r_{-}-r_{--})^2}+\frac{2 i r_{-} (r_{c} r_{+} r_{--}-r_{c} r_{-}^2+r_{+} r_{-}^2-2 r_{+} r_{-} r_{--}+r_{-}^2 r_{--}) \omega_{0}}{(r_{-}-r_{c}) (r_{+}-r_{-}) (r_{-}-r_{--})}\nonumber\\
							&		&	+\frac{1}{3} (r_{c} r_{+}-2 r_{c} r_{-}+r_{c} r_{--}+r_{+} r_{-}-2 r_{+} r_{--}+r_{-} r_{--}-1).
\label{eq:lambda_01_4DEGBdS_2}
\end{eqnarray}
In this case, the fundamental mode angular eigenvalues $\lambda_{0;1}$ are real and negative, and hence the general degree $\nu_{0;1}$ will be complex, which can be numerically evaluated from Eq.~(\ref{eq:lambda_01_4DEGBdS_2}). In fact, one of the first application of complex angular momentum techniques to atomic and molecular scattering was reported in the end of the 1970s \cite{MolecularPhysics.37.1703}, where it was evaluated the Legendre functions of the first kind of complex degree $\nu$.

\begin{figure}[p]
\centering
\includegraphics[scale=1]{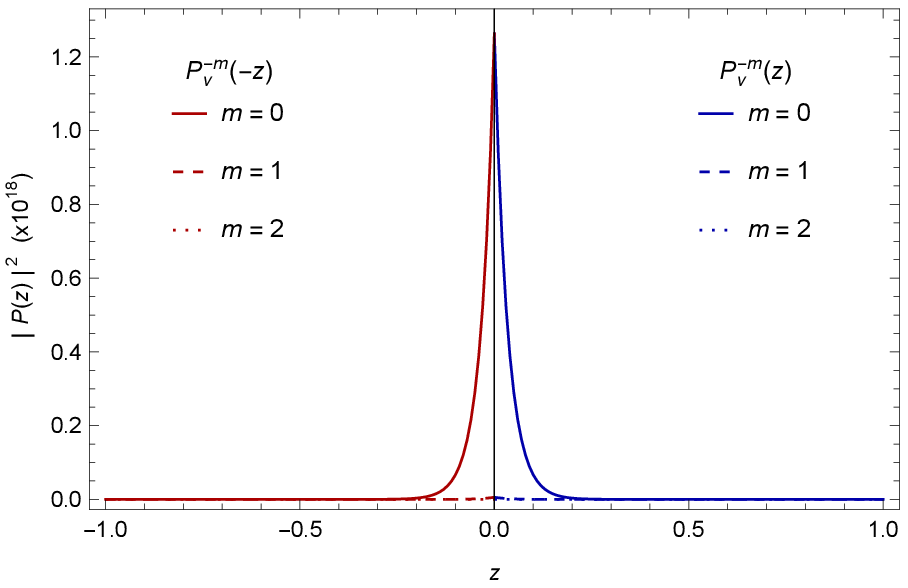}
\caption{(dS) Case 2: $a > 0$. The first three squared QBS angular wave eigenfunctions as a function of the new polar coordinate $z$; $a=0.20$, $\Lambda=0.1$, $n=0$, $M=1/2$.}
\label{fig:Fig7_4DEGBdS}
\end{figure}

In Table \ref{tab:I_4DEGBdS}, we present the fundamental mode angular eigenvalues $\lambda_{0;1}$, as well as the corresponding general degree $\nu_{0;1}$, as functions of the GB coupling constant $a$, for some values of the cosmological constant $\Lambda$. In addition, we also present the behavior of the QBS angular wave eigenfunctions $P(\theta)$ in Fig.~\ref{fig:Fig7_4DEGBdS}, as functions of the new polar coordinate $z=\cos\theta$, for some values of the magnetic quantum number $m$.  Note that these angular solutions are regular at the two boundaries $\theta=\pi$ ($z=-1$) and $\theta=0$ ($z=1$).

%
%%%%%%%%%%%%%%%%%%%%%%%%%%%%%%%%%%%%%%%%%%%%%%%%%%%%%%%%%%%%%%%%%%%%%%%%%%%%%%%%%%%%%%%%%%%%%% 4DEGBAdS
%
\section{The 4DEGB-Anti-de Sitter black hole spacetime}\label{4DEGBAdS}
For a 4DEGBAdS black hole spacetime, we will use the relation $\Lambda/3=-1/R^{2}$, where $R$ is the AdS cosmological radius, and then fix $R=1$, measure the black hole (interior and exterior) radius in units of $R$, and calculate the corresponding value of the total mass $M$ for each fixed value of the GB coupling constant in the stability region $-0.20 \leq a \leq 0.20$.

\begin{figure}[p]
\centering
\includegraphics[width=1\columnwidth]{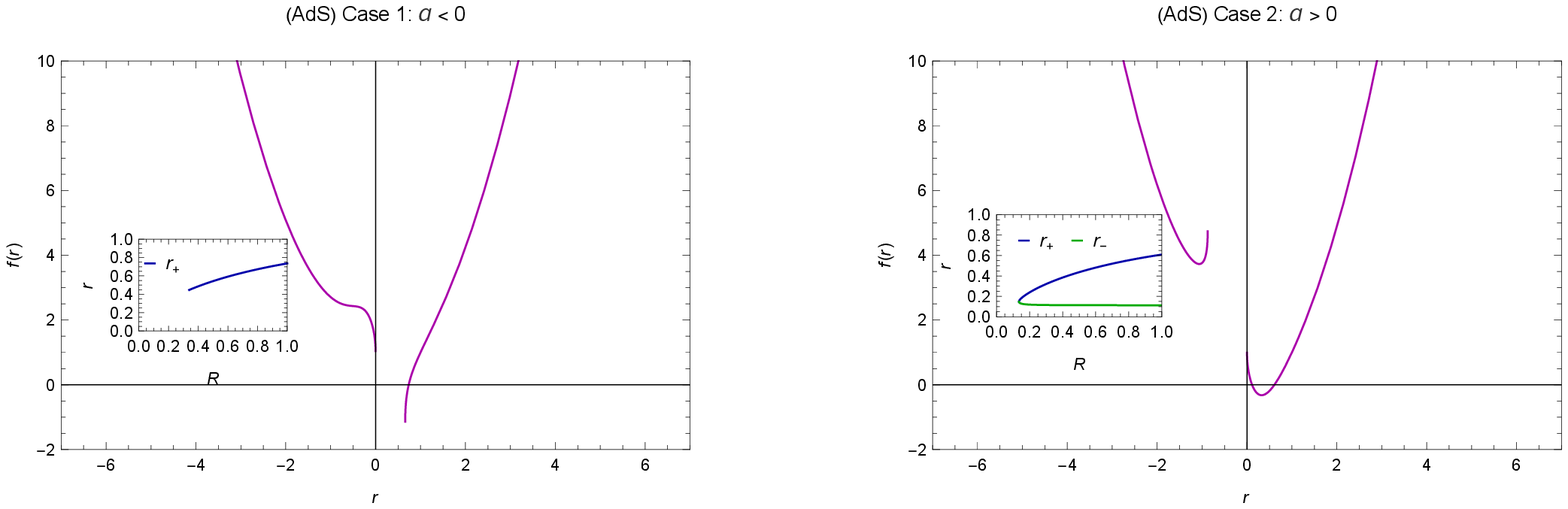}
\caption{The metric function $f(r)$ for $a=-0.20$ (left) and $a=0.20$ (right); $M=1/2$.}
\label{fig:Fig1_4DEGBAdS}
\end{figure}

We show the behavior of the metric function $f(r)$ in Fig.~\ref{fig:Fig1_4DEGBAdS}. In addition, the horizons are shown in the inlay plots as functions of the AdS cosmological radius $R$. We can see that the surface equation $f(r)=0$ has one real solution when $a < 0$, otherwise, there are two real solutions when $a > 0$.

%
%%%%%%%%%%%%%%%%%%%%%%%%%%%%%%%%%%%%%%%%%%%%%%%%%%%%%%%%%%%%%%%%%%%%%%%%%%%%%%%%%%%%%%%%%%%%%% (AdS) Case 1: a < 0
%
\subsection{(AdS) Case 1: \texorpdfstring{$a < 0$}{a < 0}}
In this case, we adopt the following parametrization for the metric function,
\begin{equation}
f(r)=\frac{1}{r^{2}}(r-r_{+}),
\label{eq:metric_function_4DEGBAdS_1}
\end{equation}
where $r_{+}$ is the (positive) ``exterior'' black hole event horizon. Then, with the metric function given by Eq.~(\ref{eq:metric_function_4DEGBAdS_1}), we will show that Eq.~(\ref{eq:radial_equation_4DEGB}) is totally convenient to study QBSs with purely ingoing boundary conditions at the exterior black hole event horizon and vanishing boundary conditions at infinity, since it is a (biconfluent) Heun-type equation \cite{Ronveaux:1995} with one finite regular singularity (associated to the exterior black hole event horizon) and one irregular singularity of rank 2 at (spatial) infinity.

Now, we also follow the steps described in the VBK approach \cite{AnnPhys.373.28,PhysRevD.104.024035}, as well as in the VB description of the biconfluent Heun functions \cite{JMathPhys.56.092501}, to obtain the analytical solution of Eq.~(\ref{eq:radial_equation_4DEGB}) in the 4DEGBAdS black hole spacetime with $a < 0$ (without the assumption of specific boundary conditions). First of all, we need to define a new radial coordinate, $x$, as
\begin{equation}
x=\frac{1}{\kappa}(r-r_{+}),
\label{eq:radial_coordinate_4DEGBAdS_1}
\end{equation}
where $\kappa(=-1^{1/4}/\omega^{1/2})$ is a (dimensionless) constant. This definition moves the singularity $r_{+}$ to the point $0$. The next step is to perform a special case of the \textit{s-homotopic transformation} $U(x) \mapsto y(x)$ given by
\begin{equation}
U(x)=x^{\frac{\alpha}{2}}\mbox{e}^{-\frac{\beta}{2}x-\frac{1}{2}x^{2}}y(x),
\label{eq:F-homotopic_4DEGBAdS_1}
\end{equation}
where
\begin{eqnarray}
\alpha	& = & -2ir_{+}^{2}\omega,\label{eq:alpha_4DEGBAdS_1}\\
\beta		& = & (2-2i)r_{+}\sqrt{2\omega}.\label{eq:beta_4DEGBAdS_1}
\end{eqnarray}
Thus, by substituting Eqs.~(\ref{eq:metric_function_4DEGBAdS_1})-(\ref{eq:beta_4DEGBAdS_1}) into Eq.~(\ref{eq:radial_equation_4DEGB}), we get
\begin{equation}
x\frac{d^{2} y(x)}{d x^{2}}+(1+\alpha-\beta x-2 x^{2})\frac{d y(x)}{d x}+\biggl\{(\gamma-\alpha-2)x-\frac{1}{2}[\delta+(1+\alpha)\beta]\frac{1}{x}\biggr\}y(x)=0,
\label{eq:radial_final_4DEGBAdS_1}
\end{equation}
where
\begin{eqnarray}
\gamma		& = & 2ir_{+}\omega,\label{eq:gamma_4DEGBAdS_1}\\
\delta		& = & \frac{1}{3}\sqrt{\frac{2}{\omega}}(1+i)(1+3\lambda-12r_{+}^{3}\omega^{2}).\label{eq:delta_4DEGBAdS_1}
\end{eqnarray}

Equation (\ref{eq:radial_final_4DEGBAdS_1}) has the (canonical) form of a biconfluent Heun equation, where $y(x) \equiv \mbox{HeunB}(\alpha,\beta,\gamma,\delta;x)$ denotes the biconfluent Heun function, which is the solution corresponding to the exponent 0 at $x=0$ and assumes the value 1 there. If $\alpha$ is not a relative integer number, then from the Fuchs-Frobenius theory it follows that the $\mbox{HeunB}(\alpha,\beta,\gamma,\delta;x)$ exists, and can be expanded as
\begin{equation}
\mbox{HeunB}(\alpha,\beta,\gamma,\delta;x)=\sum_{n \geq 0}^{\infty}c_{n}x^{n+\rho}.
\label{eq:serie_HeunB_todo_x}
\end{equation}
with
\begin{eqnarray}
\rho(\rho+\alpha) & = & 0,\nonumber\\
(\rho+1)(\rho+1+\alpha)c_{1}-\biggl\{\beta\rho+\frac{1}{2}[\delta+(1+\alpha)\beta]\biggr\}c_{0} & = & 0,
\label{eq:conditions_biconfluent_Heun}
\end{eqnarray}
where the indicial equation has two solutions, namely, $\rho=0$ and $\rho=-\alpha$. Let us focus in the $\rho=0$ solution. Thus, the $\mbox{HeunB}(\alpha,\beta,\gamma,\delta;x)$ is a entire function and given by
\begin{equation}
\mbox{HeunB}(\alpha,\beta,\gamma,\delta;x)=\sum_{n \geq 0}\frac{A_{n}}{(1+\alpha)_{n}}\frac{x^{n}}{n!}\ ,
\label{eq:Biconfluent_Heun_expansion}
\end{equation}
with
\begin{eqnarray}
A_{0}		& = &	c_{0}=1\ ,\nonumber\\
A_{1}		& = &	(1+\alpha)c_{1}=\frac{1}{2}[\delta+(1+\alpha)\beta]\ ,\nonumber\\
A_{n+2}	& = &	\biggl\{(n+1)\beta+\frac{1}{2}[\delta+(1+\alpha)\beta]\biggr\}A_{n+1}-(n+1)(n+1+\alpha)(\gamma-\alpha-2-2n)A_{n} \quad (n \geq 0),
\label{eq:recursion_Biconfluent_Heun_expansion}
\end{eqnarray}
where $(1+\alpha)_{n}=\Gamma(1+\alpha+n)/\Gamma(1+\alpha)$. On the other hand, for $\rho=-\alpha$, the solution of Eq.~(\ref{eq:radial_final_4DEGBAdS_1}) corresponding to the exponent $-\alpha$ at $x=0$ is $x^{-\alpha}\mbox{HeunB}(-\alpha,\beta,\gamma,\delta;x)$. In addition, when $\alpha$ is a negative integer number, such that $\alpha=-m$ with $m \geq 1$, it is possible to define the function $\mbox{HeunB}(-m,\beta,\gamma,\delta;x)=x^{m}\mbox{HeunB}(m,\beta,\gamma,\delta;x)$.

Therefore, the analytical solution for the radial part of the conformally coupled massless Klein-Gordon equation, in the 4DEGBAdS black hole spacetime with $a < 0$, can be written as
\begin{equation}
U_{j}(x)=x^{\frac{\alpha}{2}}\mbox{e}^{-\frac{\beta}{2}x-\frac{1}{2}x^{2}} [C_{1,j}\ y_{1,j}(x) + C_{2,j}\ y_{2,j}(x)]
\label{eq:analytical_solution_radial_4DEGBAdS_1}
\end{equation}
where $C_{1,j}$ and $C_{2,j}$ are constants to be determined, and $j=\{0,\infty\}$ labels the solution at each singular point. Thus, the pair of linearly independent solutions at $x=0$ ($r=r_{+}$) is given by
\begin{eqnarray}
y_{1,0} & = & \mbox{HeunB}(\alpha,\beta,\gamma,\delta;x),\label{eq:y10_4DEGBAdS_1}\\
y_{2,0} & = & x^{-\alpha}\mbox{HeunB}(-\alpha,\beta,\gamma,\delta;x).\label{eq:y20_4DEGBAdS_1}
\end{eqnarray}
Furthermore, the solutions of Eq.~(\ref{eq:radial_final_4DEGBAdS_1}) at infinity ($x \rightarrow \infty$) have the following asymptotic developments
\begin{eqnarray}
y_{1,\infty} & = & x^{\frac{1}{2}(\gamma-\alpha-2)},\label{eq:y1i_4DEGBAdS_1}\\
y_{2,\infty} & = & x^{-\frac{1}{2}(\gamma+\alpha+2)}\mbox{e}^{\beta x+x^{2}},\label{eq:y2i_4DEGBAdS_1}
\end{eqnarray}
where $|\arg x| \leq \pi/2 - \epsilon$.

Next, we also use the VBK approach \cite{AnnPhys.373.28,PhysRevD.104.024035} to derive the characteristic resonance equation and then find the spectrum of resonant frequencies related to quasibound states by imposing two boundary conditions on the radial solution: it should describe an ingoing wave at the exterior black hole event horizon ($r \rightarrow r_{+}$) and tend to zero far from the black hole at spatial infinity ($r \rightarrow \infty$).
%
%%%%%%%%%%%%%%%%%%%%%%%%%%%%%%%%%%%%%%%%%%%%%%%%%%%%%%%%%%%%%%%%%%%%%%%%%%%%%%%%%%%%%%%%%%%%%% Hawking radiation
%
\subsubsection{Hawking radiation}
In the limit when $r \rightarrow r_{+}$, which implies that $x \rightarrow 0$, the radial solution given by Eq.~(\ref{eq:analytical_solution_radial_4DEGBAdS_1}) behaves as
\begin{equation}
\Psi_{0}(x,t) \sim C_{1,0}\ \Psi_{{\rm in},0} + C_{2,0}\ \Psi_{{\rm out},0},
\label{eq:Hawking_4DEGBAdS_1}
\end{equation}
where
\begin{eqnarray}
\Psi_{{\rm in},0}(x>0)	& = &	\mbox{e}^{-i \omega t}x^{-\frac{i\omega}{2\kappa_{+}}},\label{eq:Hawking_in_4DEGBAdS_1}\\
\Psi_{{\rm out},0}(x>0)	& = &	\mbox{e}^{-i \omega t}x^{\frac{i\omega}{2\kappa_{+}}},\label{eq:Hawking_out_4DEGBAdS_1}
\end{eqnarray}
with
\begin{equation}
\frac{\alpha}{2}=-\frac{i\omega}{2\kappa_{+}},
\label{eq:gamma_Hawking_4DEGBAdS_1}
\end{equation}
and
\begin{equation}
\kappa_{+} \equiv \frac{1}{2r_{+}^{2}} \left.\frac{df(r)}{dr}\right|_{r=r_{+}} = \frac{1}{2r_{+}^{2}}.
\label{eq:grav_acc_4DEGBAdS_1}
\end{equation}
Therefore, these wave solutions give a thermal spectrum analogous to the spectrum of black body radiation.
%
%%%%%%%%%%%%%%%%%%%%%%%%%%%%%%%%%%%%%%%%%%%%%%%%%%%%%%%%%%%%%%%%%%%%%%%%%%%%%%%%%%%%%%%%%%%%%% Quasibound states
%
\subsubsection{Quasibound states}
In order to satisfy purely ingoing boundary conditions, from the asymptotic behavior of the radial solution at the exterior black hole event horizon described by Eq.~(\ref{eq:Hawking_4DEGBAdS_1}), we must impose that $C_{2,0}=0$.

On the other hand, in the limit when $r \rightarrow \infty$, which implies that $x \rightarrow \infty$, the radial solution given by Eq.~(\ref{eq:analytical_solution_radial_4DEGBAdS_1}) behaves as
\begin{equation}
\lim_{x \rightarrow \infty} U_{\infty}(x) \sim C_{1,\infty}\ x^{-1+\frac{\gamma}{2}}\mbox{e}^{-\frac{\beta}{2}x-\frac{1}{2}x^{2}}.
\label{eq:radial_infinity_4DEGBAdS_1}
\end{equation}
In the present case, it is easy to see that the radial solution tends to zero far from the black hole at spatial infinity, since the exponential (negative) term dominates its asymptotic behavior. However, the final asymptotic behavior of the radial solution at spatial infinity will be determined when we know the values of the parameters $\beta$ and $\gamma$, which depends on the frequencies $\omega$. In order to determine this, we use a polynomial condition for the biconfluent Heun functions to match the two asymptotic solutions of the scalar radial equation in their common overlap region. It is known that the biconfluent Heun function becomes a polynomial of degree $N$ if and only if the following two conditions are satisfied \cite{Ronveaux:1995}:
\begin{eqnarray}
\gamma-\alpha-2	& = & 2N,\label{eq:gamma-condition_biconfluent_Heun}\\
A_{N+1}					& = & 0,\label{eq:delta-condition_biconfluent_Heun}
\end{eqnarray}
where $N=0,1,2,\ldots$ is the principal quantum number. Equation (\ref{eq:gamma-condition_biconfluent_Heun}) will be called as the $\gamma$ condition, which provides the expression for the frequencies $\omega_{N}$. On the other hand, the parameter $\delta$ must be appropriately chosen so that Eqs.~(\ref{eq:conditions_biconfluent_Heun}) and (\ref{eq:recursion_Biconfluent_Heun_expansion}) are consistent, which means that it is also necessary for the parameter $\delta$ to be an eigenvalue of the biconfluent Heun equation, calculated via Eq.~(\ref{eq:delta-condition_biconfluent_Heun}). Furthermore, the parameter $\delta$ contains the separation constant $\lambda$, which indicates that we could obtain the eigenvalues of the separation constant $\lambda_{N}(\omega_{N})$, corresponding to the appropriate eigenvalues of the frequencies $\omega_{N}$, from the polynomial solution for the radial equation and then use it to show the (regular) angular behavior of massless scalar QBSs in the background under consideration.

In addition, from Eq.~(\ref{eq:recursion_Biconfluent_Heun_expansion}), more especially, when $\alpha=-N-1$ with $N \geq 0$, there exist solutions for the biconfluent Heun equation (\ref{eq:radial_final_4DEGBAdS_1}) of the form
\begin{equation}
y(x)=\sum_{n=0}^{N}A_{n}D_{\frac{1}{2}(\gamma-\alpha-2)-n}(-\sqrt{2}(x+\beta/2)),
\label{eq:HeunB_Weber-Hermite}
\end{equation}
where $D_{\nu}(z)$ are the parabolic cylinder (or Weber-Hermite) functions \cite{Erdelyi:I,Erdelyi:II,Erdelyi:III}.

Thus, by imposing the $\gamma$ condition given by Eq.~(\ref{eq:gamma-condition_biconfluent_Heun}), with the parameters $\alpha$ and $\gamma$ given by Eqs.~(\ref{eq:alpha_4DEGBAdS_1}) and (\ref{eq:gamma_4DEGBAdS_1}), respectively, we obtain a first order equation for $\omega$, whose solution is the exact spectrum of QBSs:
\begin{equation}
\omega_{N}=-\frac{i(N+1)}{2r_{+}^{2}}.
\label{eq:omega_4DEGBAdS_1}
\end{equation}

In Table \ref{tab:II_4DEGBAdS} we present the QBSs $\omega_{N}$ as functions of the exterior black hole event horizon $r_{+}$. In fact, the exterior black hole event horizon $r_{+}$ is measured in terms of the AdS cosmological radius $R$, such that it is a fixed parameter which does not depend on the GB coupling constant $a$. Henceforth, the values of the GB coupling constant $a$ are used just to calculate the black hole mass $M$.

\begin{figure}[p]
\centering
\includegraphics[width=1\columnwidth]{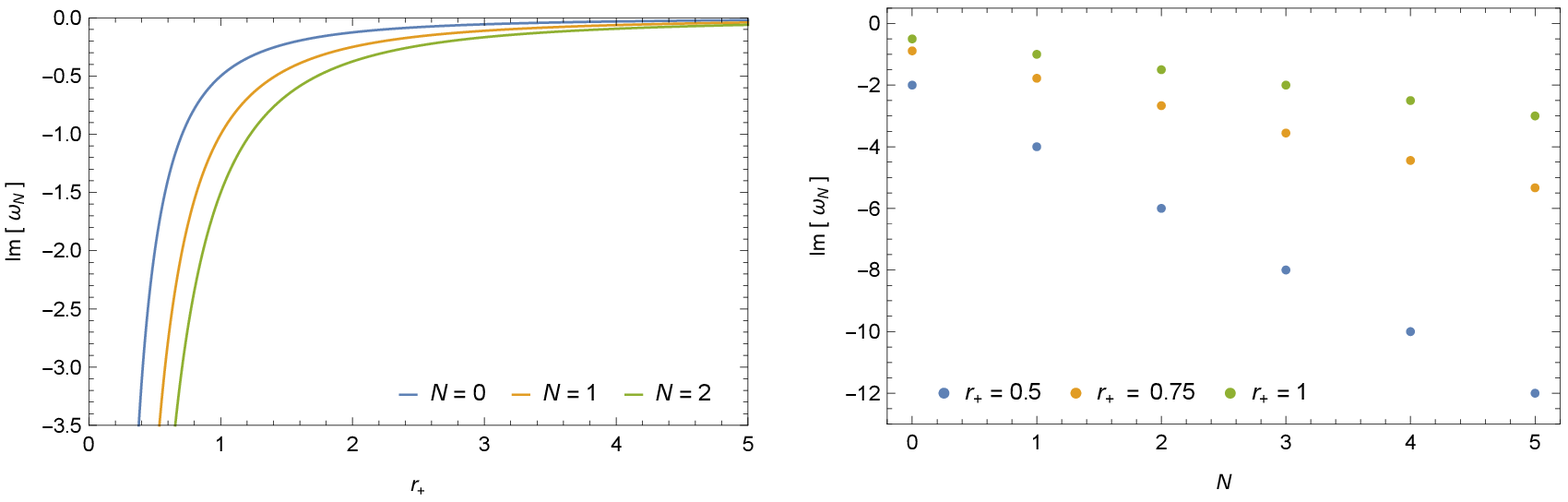}
\caption{(AdS) Case 1: $a < 0$. Massless scalar QBSs as a function of the exterior black hole event horizon (left) and principal quantum number (right).}
\label{fig:Fig2_4DEGBAdS}
\end{figure}

In addition, we also present the behavior of the QBSs $\omega_{N}$ in Fig.~\ref{fig:Fig2_4DEGBAdS}. We can conclude that the spectrum given by Eq.~(\ref{eq:omega_4DEGBAdS_1}) is physically admissible, and therefore it represents QBS frequencies for conformally coupled massless scalars in the 4DEGBdS black hole spacetime with $a < 0$. Furthermore, we can see that the decay is overdamped, since this quasispectrum is purely imaginary.

\begin{table}[t]
	\caption{(AdS) Cases 1 and 2. Massless scalar fundamental QBSs $\omega_{N}$, and the angular eigenvalue $\lambda_{0;1}$ and their corresponding general degree $\nu_{0;1}$, for some values of the exterior black hole event horizon $r_{+}$; $N=0$.}
	\label{tab:II_4DEGBAdS}
	\begin{tabular}{c|c|c|c}
		\hline\noalign{\smallskip}
		$a$							& $\omega_{0}$	& $\lambda_{0;1}$	& $\nu_{0;1}$			\\
		\noalign{\smallskip}\hline\noalign{\smallskip}
		\multicolumn{4}{c}{$r_{+}=0.5$} 			\\
		\noalign{\smallskip}\hline\noalign{\smallskip}
		$\forall a < 0$	& $-2.00i$			& $-2.33333$			& $-0.5-1.44338i$	\\
		$0.05$					& $-148.886i$		& $11.7121$				& $2.95863$				\\
		$0.10$					& $-35.4261i$		& $5.47295$				& $1.89227$				\\
		$0.15$					& $-14.9534i$		& $3.39916$				& $1.41028$				\\
		$0.20$					& $-7.96313i$		& $2.36586$				& $1.11736$				\\
		\noalign{\smallskip}\hline\noalign{\smallskip}
		\multicolumn{4}{c}{$r_{+}=0.75$} 			\\
		\noalign{\smallskip}\hline\noalign{\smallskip}
		$\forall a < 0$	& $-0.89i$			& $-1.66667$			& $-0.5-1.19024i$	\\
		$0.05$					& $-817.553i$		& $34.5226$				& $5.39683$				\\
		$0.10$					& $-202.842i$		& $16.9503$				& $3.64732$				\\
		$0.15$					& $-89.4835i$		& $11.0963$				& $2.86843$				\\
		$0.20$					& $-49.9651i$		& $8.17166$				& $2.40201$				\\
		\noalign{\smallskip}\hline\noalign{\smallskip}
		\multicolumn{4}{c}{$r_{+}=1.0$} 			\\
		\noalign{\smallskip}\hline\noalign{\smallskip}
		$\forall a < 0$	& $-0.50i$			& $-1.33333$			& $-0.5-1.04083i$	\\
		$0.05$					& $-3200.12i$		& $79.5030$				& $8.43046$				\\
		$0.10$					& $-800.113i$		& $39.5059$				& $5.80523$				\\
		$0.15$					& $-355.662i$		& $26.1754$				& $4.64056$				\\
		$0.20$					& $-200.101i$		& $19.5113$				& $3.94537$				\\
		\noalign{\smallskip}\hline\noalign{\smallskip}
		\multicolumn{4}{c}{$r_{+}=5.0$} 			\\
		\noalign{\smallskip}\hline\noalign{\smallskip}
		$\forall a < 0$	& $-0.02i$							& $-0.53333$			& $-0.5-0.53229i$	\\
		$0.05$					& $-6.760\mbox{E}7i$		& $26000.0$				& $160.746$				\\
		$0.10$					& $-1.690\mbox{E}7i$		& $13000.0$				& $113.518$				\\
		$0.15$					& $-7.511\mbox{E}7i$		& $8666.63$				& $92.5961$				\\
		$0.20$					& $-4.225\mbox{E}7i$		& $6499.96$				& $80.1239$				\\
		\noalign{\smallskip}\hline
	\end{tabular}
\end{table}

%
%%%%%%%%%%%%%%%%%%%%%%%%%%%%%%%%%%%%%%%%%%%%%%%%%%%%%%%%%%%%%%%%%%%%%%%%%%%%%%%%%%%%%%%%%%%%%% Radial wave eigenfunctions
%
\subsubsection{Radial wave eigenfunctions}
The radial wave eigenfunctions, which are related to QBSs of conformally coupled massless scalars propagating in the 4DEGBAdS black hole spacetime with $a < 0$, can be obtained by using the $\delta$ condition given by Eq.~(\ref{eq:delta-condition_biconfluent_Heun}), as well as by imposing the $\gamma$ condition given by Eq.~(\ref{eq:gamma-condition_biconfluent_Heun}) to such a parameter, as it is described in the VBK approach \cite{AnnPhys.373.28,PhysRevD.104.024035}.

As we explained before, these polynomial radial eigenfunctions are related to the appropriate determination of the eigenvalue $\delta$. Since it is calculated via Eq.~(\ref{eq:delta-condition_biconfluent_Heun}), we index its solutions by a parameter $s(=1,2,3,\ldots)$, which can be denoted by $\delta_{N;s}$. Thus, the corresponding biconfluent Heun polynomials are now denoted as $\mbox{HeunBp}_{N;s}(x)$. Therefore, the QBS radial wave eigenfunctions for conformally coupled massless scalars propagating in a 4DEGBAdS black hole spacetime with $a < 0$ are given by
\begin{equation}
U_{N;s}(x)=C_{N;s}\ x^{\frac{\alpha}{2}}\mbox{e}^{-\frac{\beta}{2}x-\frac{1}{2}x^{2}}\ \mbox{HeunBp}_{N;s}(-N-1,\beta,\gamma,\delta_{N;s};x),
\label{eq:radial_eigenfunctions_4DEGBAdS_1}
\end{equation}
where $C_{N;s}$ is a constant to be determined.

Next, we calculate the biconfluent Heun polynomials related to the fundamental and first excited modes. In the present case, we have that $\alpha=-N-1$, and hence we will write the biconfluent Heun polynomials by using Eq.~(\ref{eq:HeunB_Weber-Hermite}). The biconfluent Heun polynomial for the fundamental mode $N=0$ is given by
\begin{equation}
\mbox{HeunBp}_{0;1}(x)=A_{0}D_{0}(2i-\sqrt{2}x),
\label{eq:HeunBp_0,1}
\end{equation}
where the eigenvalue $\delta_{0;1}$ must obey
\begin{equation}
A_{1}=\frac{1}{2}[\delta+(1+\alpha)\beta]=0,
\label{eq:A_1}
\end{equation}
whose unique solution ($s=1$) is
\begin{equation}
\delta_{0;1}=-(1+\alpha)\beta.
\label{eq:delta_0,1}
\end{equation}
On the other hand, the biconfluent Heun polynomials for the first excited mode $N=1$ are given by
\begin{equation}
\mbox{HeunGp}_{1;s}(x)=A_{0}D_{1}(2i\sqrt{2}-\sqrt{2}x)+A_{1}D_{0}(2i\sqrt{2}-\sqrt{2}x),
\label{eq:HeunBp_1,s}
\end{equation}
where the eigenvalues $\delta_{1;s}$ must obey
\begin{equation}
A_{2}=0,
\label{eq:A_2}
\end{equation}
whose two solutions ($s=1,2$) are
\begin{equation}
\delta_{1;1}=-2\beta-\alpha\beta-\sqrt{-8-12\alpha-4\alpha^{2}+\beta^{2}+4\gamma+4\alpha\gamma},
\label{eq:delta_1,1}
\end{equation}
\begin{equation}
\delta_{1;2}=-2\beta-\alpha\beta+\sqrt{-8-12\alpha-4\alpha^{2}+\beta^{2}+4\gamma+4\alpha\gamma}.
\label{eq:delta_1,2}
\end{equation}
It is worth noticing that the polynomial radial eigenfunctions for the first excited mode $N=1$ are degenerate, since there exist two solutions for the eigenvalue $\delta$. Here, we choose the first solution $s=1$, and hence the eigenvalue $\delta_{1;1}$.

\begin{figure}[p]
\centering
\includegraphics[width=1\columnwidth]{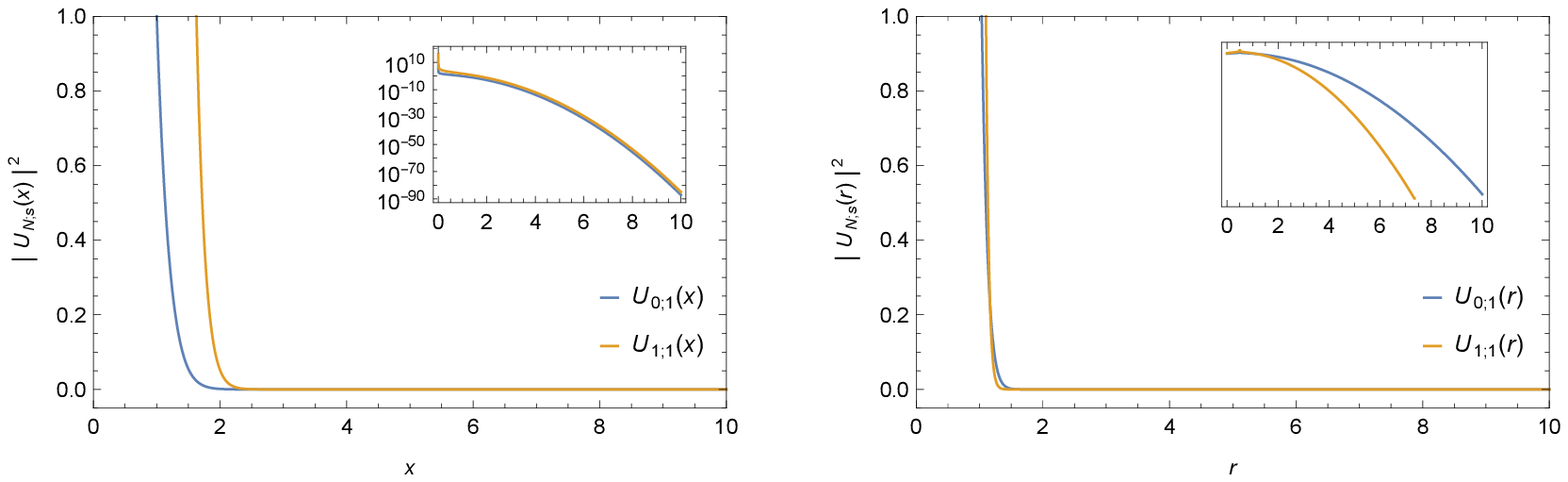}
\caption{(AdS) Case 1: $a < 0$. The first two squared QBS radial wave eigenfunctions as a function of the new radial coordinate $x$ (left) and radial coordinate $r$ (right); $a=-0.20$, $R=1$, $M=0.2125$. The units are in multiples of $C_{N;s}$.}
\label{fig:Fig3_4DEGBAdS}
\end{figure}

In Fig.~\ref{fig:Fig3_4DEGBAdS}, we plot the first two squared QBS radial wave eigenfunctions. We observe that the radial solution tends to zero at spatial infinity ($x \rightarrow \infty$) and diverges at the exterior black hole event horizon, which represents QBSs. Note that the radial wave eigenfunctions reach a maximum value at the exterior black hole event horizon ($x=0$ or $r_{+}=0.5$), and then cross this surface, as shown in the log-scale plots.

%
%%%%%%%%%%%%%%%%%%%%%%%%%%%%%%%%%%%%%%%%%%%%%%%%%%%%%%%%%%%%%%%%%%%%%%%%%%%%%%%%%%%%%%%%%%%%%% Angular wave eigenfunctions
%
\subsubsection{Angular wave eigenfunctions}
In the present case, from Eqs.~(\ref{eq:delta_4DEGBAdS_1}) and (\ref{eq:delta_0,1}), we obtain the following expression for the fundamental mode angular eigenvalues with $a < 0$:
\begin{equation}
\lambda_{0;1}=-\frac{3+r_{+}}{3r_{+}}.
\label{eq:lambda_01_4DEGBAdS_1}
\end{equation}
In this case, the fundamental mode angular eigenvalues $\lambda_{0;1}$ are real and negative, and hence the general degree $\nu_{0;1}$ will be complex, which can be numerically evaluated from Eq.~(\ref{eq:lambda_01_4DEGBAdS_1}).

\begin{figure}[p]
\centering
\includegraphics[scale=1]{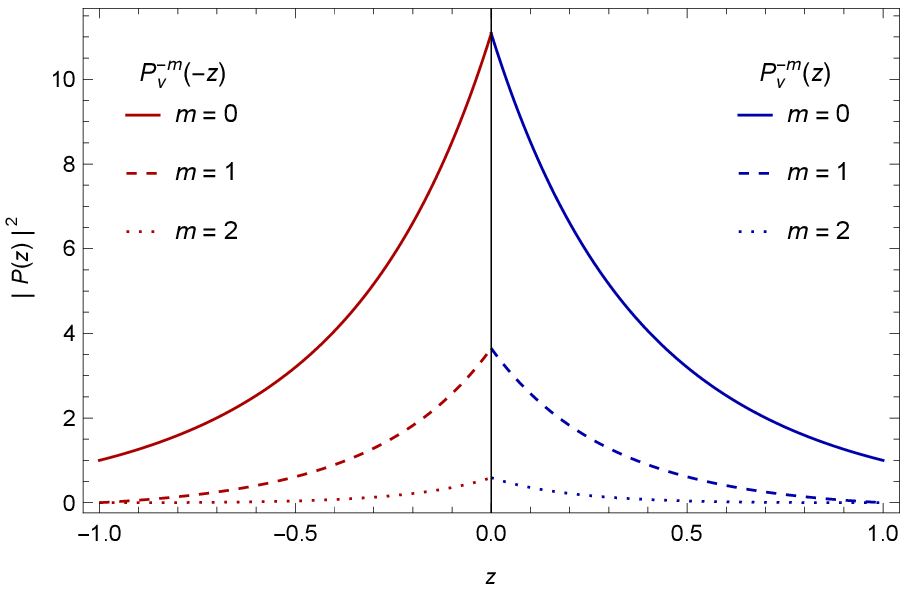}
\caption{(AdS) Case 1: $a < 0$. The first three squared QBS angular wave eigenfunctions as a function of the new polar coordinate $z$; $a=-0.20$, $R=1$, $N=0$, $M=0.2125$.}
\label{fig:Fig4_4DEGBAdS}
\end{figure}

In Table \ref{tab:II_4DEGBAdS}, we present the fundamental mode angular eigenvalues $\lambda_{0;1}$, as well as the corresponding general degree $\nu_{0;1}$, as functions of the exterior black hole event horizon $r_{+}$. In addition, we also present the behavior of the QBS angular wave eigenfunctions $P(\theta)$ in Fig.~\ref{fig:Fig4_4DEGBAdS}, as functions of the new polar coordinate $z=\cos\theta$, for some values of the magnetic quantum number $m$.  Note that these angular solutions are regular at the two boundaries $\theta=\pi$ ($z=-1$) and $\theta=0$ ($z=1$).

%
%%%%%%%%%%%%%%%%%%%%%%%%%%%%%%%%%%%%%%%%%%%%%%%%%%%%%%%%%%%%%%%%%%%%%%%%%%%%%%%%%%%%%%%%%%%%%% (AdS) Case 2: a > 0
%
\subsection{(AdS) Case 2: \texorpdfstring{$a > 0$}{a > 0}}
In this case, we adopt the following parametrization for the metric function $f(r)$,
\begin{equation}
f(r)=\frac{1}{r^{2}}(r-r_{+})(r-r_{-}),
\label{eq:metric_function_4DEGBAdS_2}
\end{equation}
where $r_{+}$ is the (positive) ``exterior'' black hole event horizon, and $r_{-}$ is the (positive) ``interior'' black hole event horizon. Then, with the metric function given by Eq.~(\ref{eq:metric_function_4DEGBAdS_2}), we will show that Eq.~(\ref{eq:radial_equation_4DEGB}) is totally suitable to study QBSs with purely ingoing boundary conditions at the exterior black hole event horizon and vanishing boundary conditions at infinity, since it is a (confluent) Heun-type equation \cite{Ronveaux:1995} with two finite regular singularities (associated to the black hole event horizons) and one irregular singularity of rank 1 at (spatial) infinity.

Now, we also follow the steps described in the VBK approach \cite{AnnPhys.373.28,PhysRevD.104.024035}, to obtain the analytical solution of Eq.~(\ref{eq:radial_equation_4DEGB}) in the 4DEGBAdS black hole spacetime with $a > 0$ (without the assumption of specific boundary conditions). First of all, we need to define a new radial coordinate, $x$, as
\begin{equation}
x=\frac{r-r_{-}}{r_{+}-r_{-}}.
\label{eq:radial_coordinate_4DEGBAdS_2}
\end{equation}
This definition move the singularities ($r_{+}$,$r_{-}$) to the points ($0$,$1$). The next step is to perform a special case of the \textit{s-homotopic transformation} $U(x) \mapsto y(x)$ given by
\begin{equation}
U(x)=x^{\frac{\beta}{2}}(x-1)^{\frac{\gamma}{2}}\mbox{e}^{\frac{\alpha}{2}x}y(x),
\label{eq:F-homotopic_4DEGBAdS_2}
\end{equation}
where
\begin{eqnarray}
\alpha	& = & -2i(r_{+}-r_{-})\omega,\label{eq:alpha_4DEGBAdS_2}\\
\beta		& = & -\frac{2ir_{-}^{2}\omega}{r_{+}-r_{-}},\label{eq:beta_4DEGBAdS_2}\\
\gamma	& = & -\frac{2ir_{+}^{2}\omega}{r_{+}-r_{-}}.\label{eq:gamma_4DEGBAdS_2}
\end{eqnarray}
Thus, by substituting Eqs.~(\ref{eq:metric_function_4DEGBAdS_2})-(\ref{eq:beta_4DEGBAdS_2}) into Eq.~(\ref{eq:radial_equation_4DEGB}), we get
\begin{equation}
\frac{d^{2} y(x)}{d x^{2}}+\biggl(\alpha+\frac{\beta+1}{x}+\frac{\gamma+1}{x-1}\biggr)\frac{d y(x)}{d x}+\biggl(\frac{\xi}{x}+\frac{\zeta}{x-1}\biggr)y(x)=0,
\label{eq:radial_final_4DEGBAdS_2}
\end{equation}
where
\begin{eqnarray}
\xi		& = &	\frac{1}{2}(\alpha-\beta-\gamma+\alpha\beta-\beta\gamma)-\eta,\label{eq:xi_4DEGBAdS_2}\\
\zeta	& = &	\frac{1}{2}(\alpha+\beta+\gamma+\alpha\gamma+\beta\gamma)+\delta+\eta,\label{eq:zeta_4DEGBAdS_2}
\end{eqnarray}
with
\begin{eqnarray}
\delta	& = & 2(r_{+}^{2}-r_{-}^{2})\omega^{2},\label{eq:delta_4DEGBAdS_2}\\
\eta		& = & \frac{2r_{-}^{3}(r_{-}-2r_{+})\omega^{2}}{(r_{+}-r_{-})^{2}}-\lambda.\label{eq:eta_4DEGBAdS_2}
\end{eqnarray}

Equation (\ref{eq:radial_final_4DEGBAdS_2}) has the (canonical) form of a confluent Heun equation, where $y(x) \equiv \mbox{HeunC}(\alpha,\beta,\gamma,\delta,\eta;x)$ denotes the confluent Heun function, which is the solution corresponding to the exponent 0 at $x=0$ and assumes the value 1 there. If $\beta$ is not a relative integer number, then this standard confluent Heun function is defined via the convergent Taylor series expansion in the disc $|x| < 1$,
\begin{equation}
\mbox{HeunC}(\alpha,\beta,\gamma,\delta,\eta;x)=\sum_{n=0}^{\infty}c_{n}(\alpha,\beta,\gamma,\delta,\eta)\ x^{n},
\label{eq:HeunC_Taylor_series}
\end{equation}
where it is assumed the normalization $\mbox{HeunC}(\alpha,\beta,\gamma,\delta,\eta;0)=1$. The coefficients $c_{n}$ are given by
\begin{equation}
P_{n}c_{n}=T_{n}c_{n-1}+X_{n}c_{n-2},
\label{eq:cn}
\end{equation}
with the initial conditions $c_{-1}=0$ and $c_{0}=1$. The expressions for $P_{n}$, $T_{n}$, and $X_{n}$ are given by
\begin{eqnarray}
P_{n} & = & 1+\frac{\beta}{n},\label{eq:Pn}\\
T_{n} & = & 1+\frac{-\alpha+\beta+\gamma-1}{n}+\frac{\eta-(-\alpha+\beta+\gamma)/2-\alpha\beta/2+\beta\gamma/2}{n^{2}},\label{eq:Tn}\\
X_{n} & = & \frac{\alpha}{n^{2}}\biggl(\frac{\delta}{\alpha}+\frac{\beta+\gamma}{2}+n-1\biggr).\label{eq:Xn}
\end{eqnarray}

Therefore, the analytical solution for the radial part of the conformally coupled massless Klein-Gordon equation, in the 4DEGBAdS black hole spacetime with $a > 0$, can be written as
\begin{equation}
U_{j}(x)=x^{\frac{\beta}{2}}(x-1)^{\frac{\gamma}{2}}\mbox{e}^{\frac{\alpha}{2}x} [C_{1,j}\ y_{1,j}(x) + C_{2,j}\ y_{2,j}(x)]
\label{eq:analytical_solution_radial_4DEGBAdS_2}
\end{equation}
where $C_{1,j}$ and $C_{2,j}$ are constants to be determined, and $j=\{0,1,\infty\}$ labels the solution at each singular point. Thus, the pair of linearly independent solutions at $x=0$ ($r=r_{-}$) is given by
\begin{eqnarray}
y_{1,0} & = & \mbox{HeunC}(\alpha,\beta,\gamma,\delta,\eta;x),\label{eq:y10_4DEGBAdS_2}\\
y_{2,0} & = & x^{-\beta}\mbox{HeunC}(\alpha,-\beta,\gamma,\delta,\eta;x),\label{eq:y20_4DEGBAdS_2}
\end{eqnarray}
where the parameters $\alpha$, $\beta$, $\gamma$, $\delta$, and $\eta$ are given by Eqs.~(\ref{eq:alpha_4DEGBAdS_2}), (\ref{eq:beta_4DEGBAdS_2}), (\ref{eq:gamma_4DEGBAdS_2}), (\ref{eq:delta_4DEGBAdS_2}), and (\ref{eq:eta_4DEGBAdS_2}), respectively. Furthermore, by following the same steps, the pair of linearly independent solutions at $x=1$ ($r=r_{+}$) is given by
\begin{eqnarray}
y_{1,1}(x \rightarrow 1-x) & = & \mbox{HeunC}(\alpha,\beta,\gamma,\delta,\eta;1-x),\label{eq:y11_4DEGBAdS_2}\\
y_{2,1}(x \rightarrow 1-x) & = & (1-x)^{-\beta}\mbox{HeunC}(\alpha,-\beta,\gamma,\delta,\eta;1-x),\label{eq:y21_4DEGBAdS_2}
\end{eqnarray}
where the new parameters $\alpha$, $\beta$, $\gamma$, $\delta$, and $\eta$ are given by
\begin{eqnarray}
\alpha	& = & 2i(r_{+}-r_{-})\omega,\label{eq:alpha1_4DEGBAdS_2}\\
\beta		& = & -\frac{2ir_{+}^{2}\omega}{r_{+}-r_{-}},\label{eq:beta1_4DEGBAdS_2}\\
\gamma	& = & -\frac{2ir_{-}^{2}\omega}{r_{+}-r_{-}},\label{eq:gamma1_4DEGBAdS_2}\\
\delta	& = & -2(r_{+}^{2}-r_{-}^{2})\omega^{2},\label{eq:delta1_4DEGBAdS_2}\\
\eta		& = & \frac{2r_{+}^{3}(r_{+}-2r_{-})\omega^{2}}{(r_{+}-r_{-})^{2}}-\lambda.\label{eq:eta1_4DEGBAdS_2}
\end{eqnarray}
Finally, the solutions of Eq.~(\ref{eq:radial_final_4DEGBAdS_2}) at infinity ($x \rightarrow \infty$) have the following asymptotic developments
\begin{eqnarray}
y_{1,\infty} & = & x^{-(\frac{\beta+\gamma+2}{2}+\frac{\delta}{\alpha})},\label{eq:y1i_4DEGBAdS_2}\\
y_{2,\infty} & = & x^{-(\frac{\beta+\gamma+2}{2}-\frac{\delta}{\alpha})}\mbox{e}^{- \alpha x}.\label{eq:y2i_4DEGBAdS_2}
\end{eqnarray}

Next, we also use the VBK approach \cite{AnnPhys.373.28,PhysRevD.104.024035} to derive the characteristic resonance equation and then find the spectrum of resonant frequencies related to quasibound states by imposing two boundary conditions on the radial solution: it should describe an ingoing wave at the exterior black hole event horizon ($r \rightarrow r_{+}$) and tend to zero far from the black hole at spatial infinity ($r \rightarrow \infty$).
%
%%%%%%%%%%%%%%%%%%%%%%%%%%%%%%%%%%%%%%%%%%%%%%%%%%%%%%%%%%%%%%%%%%%%%%%%%%%%%%%%%%%%%%%%%%%%%% Hawking radiation
%
\subsubsection{Hawking radiation}
In the limit when $r \rightarrow r_{+}$, which implies that $x \rightarrow 1$, the radial solution given by Eq.~(\ref{eq:analytical_solution_radial_4DEGBAdS_2}) behaves as
\begin{equation}
\Psi_{1}(x,t) \sim C_{1,1}\ \Psi_{{\rm in},1} + C_{2,1}\ \Psi_{{\rm out},1},
\label{eq:Hawking_4DEGBAdS_2}
\end{equation}
where
\begin{eqnarray}
\Psi_{{\rm in},1}(x>1)	& = &	\mbox{e}^{-i \omega t}x^{-\frac{i\omega}{2\kappa_{+}}},\label{eq:Hawking_in_4DEGBAdS_2}\\
\Psi_{{\rm out},1}(x>1)	& = &	\mbox{e}^{-i \omega t}x^{\frac{i\omega}{2\kappa_{+}}},\label{eq:Hawking_out_4DEGBAdS_2}
\end{eqnarray}
with
\begin{equation}
\frac{\beta}{2}=-\frac{i\omega}{2\kappa_{+}},
\label{eq:gamma_Hawking_4DEGBAdS_2}
\end{equation}
and
\begin{equation}
\kappa_{+} \equiv \frac{1}{2r_{+}^{2}} \left.\frac{df(r)}{dr}\right|_{r=r_{+}} = \frac{r_{+}-r_{-}}{2r_{+}^{2}}.
\label{eq:grav_acc_4DEGBAdS_2}
\end{equation}
Therefore, these wave solutions give a thermal spectrum analogous to the spectrum of black body radiation.
%
%%%%%%%%%%%%%%%%%%%%%%%%%%%%%%%%%%%%%%%%%%%%%%%%%%%%%%%%%%%%%%%%%%%%%%%%%%%%%%%%%%%%%%%%%%%%%% Quasibound states
%
\subsubsection{Quasibound states}
In order to satisfy purely ingoing boundary conditions, from the asymptotic behavior of the radial solution at the exterior black hole event horizon described by Eq.~(\ref{eq:Hawking_4DEGBAdS_2}), we must impose that $C_{2,1}=0$.

On the other hand, in the limit when $r \rightarrow \infty$, which implies that $x \rightarrow \infty$, the radial solution given by Eq.~(\ref{eq:analytical_solution_radial_4DEGBAdS_2}) behaves as
\begin{equation}
\lim_{x \rightarrow \infty} U_{\infty}(x) \sim C_{1,\infty}\ x^{-1}\ x^{-p}\ \mbox{e}^{-q x},
\label{eq:radial_infinity_4DEGBAdS_2}
\end{equation}
where
\begin{eqnarray}
p & = & \frac{i(r_{+}^{2}-r_{-}^{2})\omega}{r_{+}-r_{-}},\label{eq:p_infinity_4DEGBAdS_2}\\
q & = & i(r_{+}-r_{-})\omega.\label{eq:q_infinity_4DEGBAdS_2}
\end{eqnarray}

In the present case, when $\omega=-i\omega_{I}$, it is easy to see that the radial solution tends to zero far from the black hole at spatial infinity, since $p > 0$ and $q > 0$. Thus, in order to determine the frequencies $\omega$, we use a polynomial condition for the confluent Heun functions to match the two asymptotic solutions of the scalar radial equation in their common overlap region. It is known that the confluent Heun function becomes a polynomial of degree $N$ if and only if the following two conditions are satisfied \cite{Ronveaux:1995}:
\begin{eqnarray}
\frac{\delta}{\alpha}+\frac{\beta+\gamma}{2}+N+1	& = & 0,\label{eq:delta-condition_confluent_Heun}\\
\Delta_{N+1}(\xi)																	& = & 0,\label{eq:Delta-condition_confluent_Heun}
\end{eqnarray}
where $N=0,1,2,\ldots$ is the principal quantum number. Equation (\ref{eq:delta-condition_confluent_Heun}) will be called as the $\delta$ condition, which provides the expression for the frequencies $\omega_{N}$. On the other hand, the parameter $\xi$ must be appropriately chosen so that Eqs.~(\ref{eq:cn})-(\ref{eq:Xn}) are consistent, which means that it is also necessary for the parameter $\xi$ to be an eigenvalue of the confluent Heun equation, calculated via Eq.~(\ref{eq:Delta-condition_confluent_Heun}). Furthermore, the parameter $\xi$, that is, the parameter $\eta$, contains the separation constant $\lambda$, which indicates that we could obtain the eigenvalues of the separation constant $\lambda_{N}(\omega_{N})$, corresponding to the appropriate eigenvalues of the frequencies $\omega_{N}$, from the polynomial solution for the radial equation and then use it to show the (regular) angular behavior of massless scalar QBSs in the background under consideration. Note that the $\delta$ condition is equivalent to $X_{N+2}=0$ in Eq.~(\ref{eq:Xn}), and the $\Delta$ condition is equivalent to the requirement $c_{N+1}=0$ in Eq.~(\ref{eq:cn}).

Thus, by imposing the $\delta$ condition given by Eq.~(\ref{eq:delta-condition_confluent_Heun}), we obtain a first order equation for $\omega$, whose solution is the exact spectrum of QBSs:
\begin{equation}
\omega_{N}=-\frac{i(N+1)(r_{+}-r_{-})}{2r_{-}^{2}}.
\label{eq:omega_4DEGBAdS_2}
\end{equation}

In Table \ref{tab:II_4DEGBAdS} we present the QBSs $\omega_{N}$ as functions of the exterior black hole event horizon $r_{+}$ for some values of the GB coupling constant $a$. In fact, the exterior black hole event horizon $r_{+}$ is measured in terms of the AdS cosmological radius $R$, such that it is a fixed parameter which does not depend on the GB coupling constant $a$. On the other hand, the interior black hole event horizon $r_{-}$ depends on the GB coupling constant $a$.

\begin{figure}[p]
\centering
\includegraphics[width=1\columnwidth]{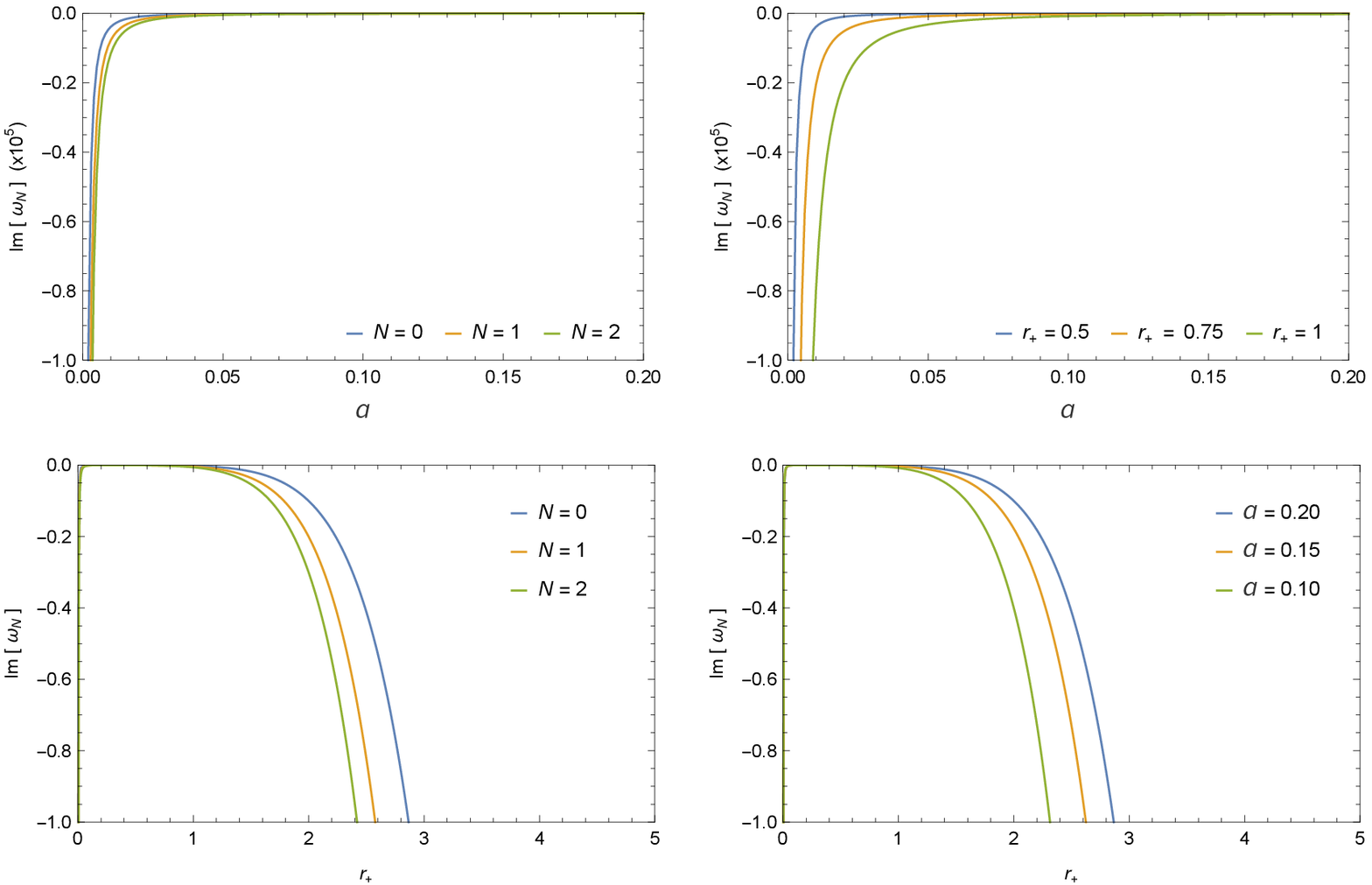}
\caption{(AdS) Case 2: $a > 0$. Top panel: Massless scalar QBSs as a function of the GB coupling constant for $r_{+}=0.5$ (left) and $N=0$ (right). Bottom panel: Massless scalar QBSs as a function of the exterior black hole event horizon for $a=0.20$ (left) and $N=0$ (right).}
\label{fig:Fig5_4DEGBAdS}
\end{figure}

In addition, we also present the behavior of the QBSs $\omega_{N}$ in Fig.~\ref{fig:Fig5_4DEGBAdS}. We can conclude that the spectrum given by Eq.~(\ref{eq:omega_4DEGBAdS_2}) is physically admissible, and therefore it represents QBS frequencies for conformally coupled massless scalars in the 4DEGBdS black hole spacetime with $a < 0$. Furthermore, we can see that the decay is overdamped, since this quasispectrum is purely imaginary.

%
%%%%%%%%%%%%%%%%%%%%%%%%%%%%%%%%%%%%%%%%%%%%%%%%%%%%%%%%%%%%%%%%%%%%%%%%%%%%%%%%%%%%%%%%%%%%%% Radial wave eigenfunctions
%
\subsubsection{Radial wave eigenfunctions}
The radial wave eigenfunctions, which are related to QBSs of conformally coupled massless scalars propagating in the 4DEGBAdS black hole spacetime with $a > 0$, can be obtained by using the $\delta$ condition given by Eq.~(\ref{eq:delta-condition_confluent_Heun}), as well as by imposing the $\Delta$ condition given by Eq.~(\ref{eq:Delta-condition_confluent_Heun}) to parameter $\xi$, as it is described in the VBK approach \cite{AnnPhys.373.28,PhysRevD.104.024035}.

As we explained before, these polynomial radial eigenfunctions are related to the appropriate determination of the eigenvalue $\xi$. Since it is calculated via Eq.~(\ref{eq:Delta-condition_confluent_Heun}), we index its solutions by a parameter $s(=1,2,3,\ldots)$, which can be conveniently denoted by $\xi_{N;s}$. Thus, the corresponding confluent Heun polynomials are now denoted as $\mbox{HeunCp}_{N;s}(x)$. Therefore, the QBS radial wave eigenfunctions for conformally coupled massless scalars propagating in a 4DEGBAdS black hole spacetime with $a > 0$ are given by
\begin{equation}
U_{N;s}(x)=C_{N;s}\ x^{\frac{\beta}{2}}(x-1)^{\frac{\gamma}{2}}\mbox{e}^{\frac{\alpha}{2}x}\ \mbox{HeunCp}_{N;s}(\alpha,\beta,\gamma,\delta,\eta;x),
\label{eq:radial_eigenfunctions_4DEGBAdS_2}
\end{equation}
where $C_{N;s}$ is a constant to be determined.

Next, we calculate the confluent Heun polynomials related to the fundamental and first excited modes. The confluent Heun polynomial for the fundamental mode $N=0$ is given by
\begin{equation}
\mbox{HeunCp}_{0;1}(x)=1,
\label{eq:HeunCp_0,1}
\end{equation}
where the eigenvalue $\xi_{0;1}$ must obey
\begin{equation}
c_{1}=\frac{T_{1}}{P_{1}}=0,
\label{eq:c_1_HeunCp_0,1}
\end{equation}
whose unique solution ($s=1$) is
\begin{equation}
\xi_{0;1}=0.
\label{eq:xi_0,1}
\end{equation}
On the other hand, the confluent Heun polynomials for the first excited mode $N=1$ are given by
\begin{equation}
\mbox{HeunCp}_{1;s}(x)=c_{0}+c_{1}x=1-\frac{\xi_{1;s}}{1+\beta}x,
\label{eq:HeunCp_1,s}
\end{equation}
where the eigenvalues $\xi_{1;s}$ must obey
\begin{equation}
c_{2}=\frac{1}{P_{2}}(T_{2}c_{1}+X_{2}c_{0})=0,
\label{eq:c_2_HeunCp_1,s}
\end{equation}
whose two solutions ($s=1,2$) are
\begin{equation}
\xi_{1;1}=\frac{-(\alpha-2-\beta-\gamma) + \sqrt{\Delta}}{2},
\label{eq:xi_1,1}
\end{equation}
\begin{equation}
\xi_{1;2}=\frac{-(\alpha-2-\beta-\gamma) - \sqrt{\Delta}}{2},
\label{eq:xi_1,2}
\end{equation}
with $\Delta=(\alpha-2-\beta-\gamma)^{2}+4\alpha(1+\beta)$. It is worth noticing that the polynomial radial eigenfunctions for the first excited mode $N=1$ are degenerate, since there exist two solutions for the eigenvalue $\xi$. Here, we choose the first solution $s=1$, and hence the eigenvalue $\xi_{1;1}$.

\begin{figure}[p]
\centering
\includegraphics[width=1\columnwidth]{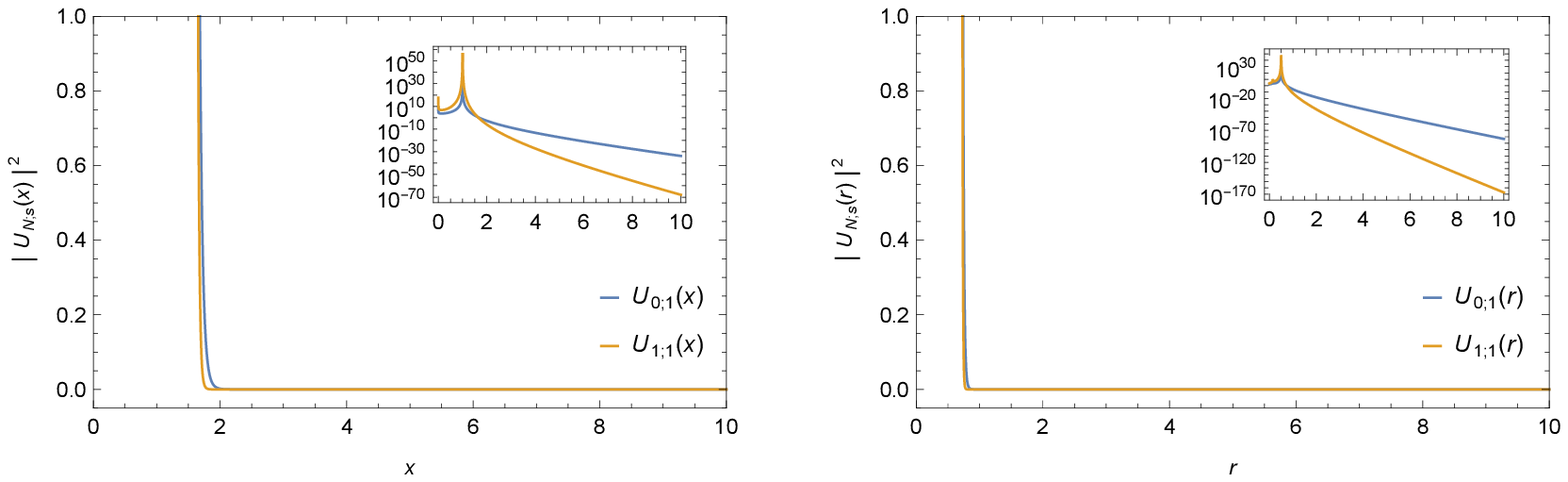}
\caption{(AdS) Case 2: $a > 0$. The first two squared QBS radial wave eigenfunctions as a function of the new radial coordinate $x$ (left) and radial coordinate $r$ (right); $a=0.20$, $R=1$, $M=0.2125$. The units are in multiples of $C_{N;s}$.}
\label{fig:Fig6_4DEGBAdS}
\end{figure}

In Fig.~\ref{fig:Fig6_4DEGBAdS}, we plot the first two squared QBS radial wave eigenfunctions. We observe that the radial solution tends to zero at spatial infinity ($x \rightarrow \infty$) and diverges at the exterior black hole event horizon, which represents QBSs. Note that the radial wave eigenfunctions reach a maximum value at the exterior black hole event horizon ($x=0$ or $r_{+}=0.5$), and then cross this surface, as shown in the log-scale plots.

%
%%%%%%%%%%%%%%%%%%%%%%%%%%%%%%%%%%%%%%%%%%%%%%%%%%%%%%%%%%%%%%%%%%%%%%%%%%%%%%%%%%%%%%%%%%%%%% Angular wave eigenfunctions
%
\subsubsection{Angular wave eigenfunctions}
In the present case, from Eqs.~(\ref{eq:xi_4DEGBAdS_2}) and (\ref{eq:xi_0,1}), we obtain the following expression for the fundamental mode angular eigenvalues with $a > 0$:
\begin{equation}
\lambda_{0;1}=-\frac{2i[r_{+}^{3}+2r_{+}r_{-}^{2}-r_{-}^{3}-2r_{+}^{2}r_{-}(1+ir_{-}\omega_{0})]\omega_{0}}{(r_{+}-r_{-})^{2}}.
\label{eq:lambda_01_4DEGBAdS_2}
\end{equation}
In this case, the fundamental mode angular eigenvalues $\lambda_{0;1}$ are real and positive, and hence the general degree $\nu_{0;1}$ will be real, which can be numerically evaluated from Eq.~(\ref{eq:lambda_01_4DEGBAdS_2}).

\begin{figure}[p]
\centering
\includegraphics[scale=1]{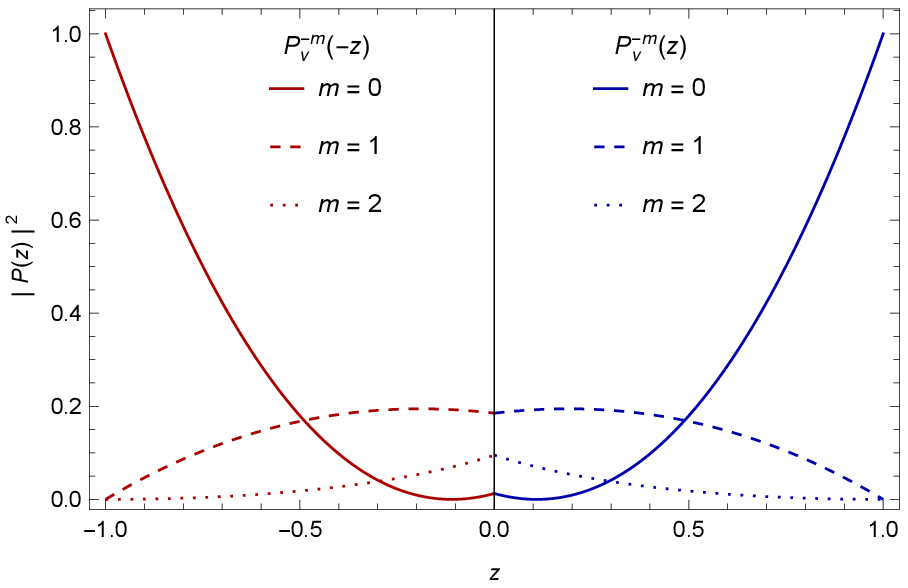}
\caption{(AdS) Case 2: $a > 0$. The first three squared QBS angular wave eigenfunctions as a function of the new polar coordinate $z$; $a=0.20$, $R=1$, $N=0$, $M=0.2125$.}
\label{fig:Fig7_4DEGBAdS}
\end{figure}

In Table \ref{tab:II_4DEGBAdS}, we present the fundamental mode angular eigenvalues $\lambda_{0;1}$, as well as the corresponding general degree $\nu_{0;1}$, as functions of the exterior black hole event horizon $r_{+}$. In addition, we also present the behavior of the QBS angular wave eigenfunctions $P(\theta)$ in Fig.~\ref{fig:Fig7_4DEGBAdS}, as functions of the new polar coordinate $z=\cos\theta$, for some values of the magnetic quantum number $m$.  Note that these angular solutions are regular at the two boundaries $\theta=\pi$ ($z=-1$) and $\theta=0$ ($z=1$).

%
%%%%%%%%%%%%%%%%%%%%%%%%%%%%%%%%%%%%%%%%%%%%%%%%%%%%%%%%%%%%%%%%%%%%%%%%%%%%%%%%%%%%%%%%%%%%%% Final remarks
%
\section{Final remarks}\label{Conclusions}
In this work, we obtained the exact analytical general solutions of the conformally coupled massless Klein-Gordon equation in the 4DEGB de Sitter and Anti-de Sitter black hole spacetimes. The general solution of the angular part is given in terms of the associated Legendre functions with arbitrary degree, while the general solutions for the radial part are given in terms of the Heun functions.

On the radial solutions, we imposed the two boundary conditions related to the quasibound state phenomena. Near the exterior event horizon, the radial solutions describe the Hawking radiation spectrum. In particular, the ingoing waves reach a maximum value and then crosses into the black hole. On the other hand, far from the black hole at asymptotic (spatial) infinity, the radial solutions tend to zero, that is, the probability of finding any scalar particles in the spatial infinity is null.

The spectrum of quasibound states for massless scalar fields was obtained by using the polynomial condition of the Heun functions. In fact, that is a new (analytical) approach developed by Vieira, Bezerra, and Kokkotas \cite{PhysRevD.104.024035,AnnPhys.373.28}.

Finally, we can discuss the stability of the systems. All the systems are stables, and present an over-damped motion, since their resonant frequencies is purely imaginary and negative. In addition, it is worth pointing out that these quasibound state frequencies were obtained directly from the Heun functions by using a polynomial condition, and, to our knowledge, there is no similar result in the literature for the backgrounds under consideration.

We hope that our results, which describe an unquestionably phenomenon associated with purely quantum effects in gravity, may be used to fit some astrophysical data in the near future and hence shed some light on the physics of black holes, and alternative/modified theories of gravity as well.
%
%%%%%%%%%%%%%%%%%%%%%%%%%%%%%%%%%%%%%%%%%%%%%%%%%%%%%%%%%%%%%%%%%%%%%%%%%%%%%%%%%%%%%%%%%%%%%% Author contributions
%
%\section*{Author contributions}
%H. S. Vieira: Conceptualization, investigation, methodology, software, writing.
%
%K. D. Kokkotas: Project administration, supervision, validation.
%
%%%%%%%%%%%%%%%%%%%%%%%%%%%%%%%%%%%%%%%%%%%%%%%%%%%%%%%%%%%%%%%%%%%%%%%%%%%%%%%%%%%%%%%%%%%%%% Data availability
%
\section*{Data availability}
The data that support the findings of this study are available from the corresponding author upon reasonable request.
%
%%%%%%%%%%%%%%%%%%%%%%%%%%%%%%%%%%%%%%%%%%%%%%%%%%%%%%%%%%%%%%%%%%%%%%%%%%%%%%%%%%%%%%%%%%%%%% Acknowledgments
%
\begin{acknowledgments}
H.S.V. is funded by the Alexander von Humboldt-Stiftung/Foundation (Grant No. 1209836). This study was financed in part by the Coordena\c c\~{a}o de Aperfei\c coamento de Pessoal de N\'{i}vel Superior - Brasil (CAPES) - Finance Code 001. V.B.B. is partially supported by the Conselho Nacional de Desenvolvimento Cient\'{i}fico e Tecnol\'{o}gico (CNPq) through the Research Project No. 307211/2020-7. C.R.M. is partially supported by the CNPq through the Research Project No. 308168/2021-6, and by the Funda\c{c}\~{a}o Cearense de Apoio ao Desenvolvimento Cient\'{\i}fico e Tecnol\'{o}gico (FUNCAP) under the grant PRONEM PNE-0112-00085.01.00/16. M.S.C is partially supported by the CNPq through the Research Project No. 315926/2021-0.
\end{acknowledgments}
%
%%%%%%%%%%%%%%%%%%%%%%%%%%%%%%%%%%%%%%%%%%%%%%%%%%%%%%%%%%%%%%%%%%%%%%%%%%%%%%%%%%%%%%%%%%%%%% thebibliography
%

%
%%%%%%%%%%%%%%%%%%%%%%%%%%%%%%%%%%%%%%%%%%%%%%%%%%%%%%%%%%%%%%%%%%%%%%%%%%%%%%%%%%%%%%%%%%%%%%
%
\end{document}